\documentclass[]{aastex63}
\usepackage{graphicx}
\usepackage{CJK}

\accepted{\today}
\submitjournal{AJ}

\begin{document}
\begin{CJK*}{UTF8}{gbsn}

\defcitealias{2024AJ....167..123L}{Paper I}

\title{Metallicity Distribution in M31 and M33 Based on the Tip-RGB Near-infrared Color}

\correspondingauthor{Biwei Jiang}
\email{bjiang@bnu.edu.cn}

\author[0009-0001-5020-4269]{Ying Li (李颖)}
\affiliation{Institute for Frontiers in Astronomy and Astrophysics, Beijing Normal University, Beijing 102206, China}
\affiliation{School of Physics and Astronomy, Beijing Normal University, Beijing 100875, China}

\author[0000-0003-3168-2617]{Biwei Jiang (姜碧沩)}
\affiliation{Institute for Frontiers in Astronomy and Astrophysics, Beijing Normal University, Beijing 102206, China}
\affiliation{School of Physics and Astronomy, Beijing Normal University, Beijing 100875, China}

\author[0000-0003-1218-8699]{Yi Ren (任逸)}
\affiliation{College of Physics and Electronic Engineering, Qilu Normal University, Jinan 250200, China}

\begin{abstract}
This study investigates the metallicity distribution in M31 and M33 by using the near-infrared color index $J-K$ of tip-red-giant-branch (TRGB) of the member stars from \cite{2021ApJ...907...18R} after removing the foreground dwarf stars by the near-infrared $J-H/H-K$ diagram as well as the Gaia astrometric measurements.
We employ the Voronoi binning technique to divide the galaxy into sub-regions, the PN method to determine the TRGB position in the $J-K$/$K$ diagram, and the bootstrap method to estimate the uncertainties.
The TRGB positions are calculated to be $J-K = 1.195 \pm 0.002$ and $1.100 \pm 0.003$, and $K = 17.615 \pm 0.007$ and $18.185 \pm 0.053$ for M31 and M33 respectively as an entity, consistent with previous results.
The $J-K$ gradient of M31 is found to be $-$0.0055 kpc$^{-1}$ at $R_{\rm GC}=(0, 24)$ kpc and $-$0.0002 kpc$^{-1}$ at $R_{\rm GC}=(24, 150)$ kpc. Notably, two dust rings are evident at 11.5 kpc and 14.5 kpc, in agreement with previous studies.
The $J-K$ of M33 is analyzed in four directions and generally shows a trend of gradually decreasing from the inside to the outside, however, it shows an increasing trend of 0.022 kpc$^{-1}$ in the inner 0$-$2 kpc in the west.
Through the color$-$metallicity relation of TRGB, the metallicity gradient of M31 turns to be $-0.040 \pm 0.0012$ dex kpc$^{-1}$ with $R_{\rm GC}<30$ kpc and $-0.001 \pm 0.0002$ dex kpc$^{-1}$ with $R_{\rm GC}>30$ kpc, and for M33, $-$0.269 $\pm$ 0.0206 dex kpc$^{-1}$ with $R_{\rm GC}<9$ kpc.
\end{abstract}

\keywords{\href{http://astrothesaurus.org/uat/1371}{Red giant tip (1371)}; \href{http://astrothesaurus.org/uat/39}{M31 (39)};
\href{http://astrothesaurus.org/uat/1712}{M33 (1712)}; \href{http://astrothesaurus.org/uat/1031}{Metallicity (1031)}
}

\section{Introduction} \label{sec:introduction}

Metallicity gradients play a vital role in understanding galaxy evolution. They provide evidence of how galaxies change through star formation, gas accretion and mergers, revealing the redistribution of metals within galaxies.
Stars synthesize heavier elements and eject them into the interstellar medium so that metallicity gradients can indicate past star formation rates and gas flows. Galaxy mergers and interactions can disrupt these gradients, flattening or steepening them, while accretion of metal-poor gas from the surroundings can dilute central metallicities, all of which offer clues about a galaxy's dynamic history.

M31, also known as the Andromeda Galaxy, is a massive spiral galaxy located approximately 770 kpc away. As the dominant member of the Local Group, it serves as a key target for studying galaxy formation and evolution.
Based on the space structure, the study of metallicity gradient can be organized into disk and halo.
Extensive research has focused on the disk.
\cite{2015AJ....150..189G} study the metallicity gradient of old red giant branch (RGB) stars, and obtain a clear gradient of $-0.020 \pm 0.004$ dex kpc$^{-1}$ with galactocentric distance ($R_{\rm GC}$) from 4 kpc to 20 kpc.
Similarly, \cite{2021AJ....162...45E} get [Fe/H] and [$\alpha$/Fe] for individual RGB stars from low-(R˜3000) and moderate-(R˜6000) resolution Keck/DEIMOS spectra, and measure a [Fe/H] gradient of $-0.018 \pm 0.003$ dex kpc$^{-1}$ with $R_{\rm GC} <$ 33 kpc.
In addition to RGB stars, other objects have also traced the disk's metallicity distribution.
\cite{2013ApJ...777...35L} analyzed the data from the Pan-STARRS1 PAndromeda project and identified 17 Cepheids distributed with $R_{\rm GC}$ range of 10$-$16 kpc. Within 15 kpc, the metallicity is lower than the solar but similar to the metallicity of the H II regions in M31, which leads to the metallicity gradient of the disk being $-0.008 \pm 0.004$ dex kpc$^{-1}$. The result from the planetary nebulae (PNe), $-0.011 \pm 0.004$ dex kpc$^{-1}$, is consistent \citep{2012ApJ...753...12K}.
\cite{2022ApJ...932...29L} use the low-resolution LAMOST and Keck spectra of blue supergiants distributed over the disk to obtain the metallicity gradient of M31 being $-0.018$ dex kpc$^{-1}$, with the metallicity at the center of 0.30 $\pm$ 0.09 dex.
Moreover, the emission lines in the spectrum can be used to measure the metal abundance gradient in the H II regions, especially to estimate the metal abundance based on the emission lines of oxygen ions. \cite{2012MNRAS.427.1463Z} find that the [O/H] gradient in M31 is $-0.023 \pm 0.002$ dex kpc$^{-1}$.
\cite{2016AJ....152...45C} use the old clusters in the inner disk (0$-$30 kpc) of M31 and find a clear metallicity gradient of $-0.038 \pm 0.023$ dex kpc$^{-1}$.
These works certify there is a negative metallicity gradient in the M31 disk, and the range is from $-0.008$ dex kpc$^{-1}$ to $-0.038$ dex kpc$^{-1}$.

Moving to the M31 halo, \cite{1991ApJ...370..495H} use six absorption features of 150 globular clusters to derive a weak metallicity gradient as a function of projected radius. They gave the mean metallicity of [Fe/H]=$-$1.2, which is higher than that of the Galaxy.
\cite{2000AJ....119..727B} give a bimodality of globular clusters in M31, which results in the mean of the two groups being [Fe/H]=$-$1.4 and $-$0.6 respectively. They suggested that the two groups are similar to Galactic `halo' and `bulge/disk'.

\cite{2014ApJ...780..128I} present an analysis of the large-scale structure of the M31 halo, and observed a substantial metallicity gradient, which declines from [Fe/H]=$-$0.7 at R=30 kpc to $-$1.5 at R=150 kpc, corresponding to a gradient of $-0.007$ dex kpc$^{-1}$.
Similarly, \cite{2024MNRAS.tmp.2458O} use RGB stars to investigate the metallicity distribution of the stellar halo without uncertainties due to foreground contamination. Their work also revealed a negative metallicity gradient, $-0.0044 \pm 0.0004$ dex kpc$^{-1}$, with the largest $R_{\rm GC}$ exceeding 100 kpc.
Additionally, \cite{2001AJ....121.2557D} determined the distribution of metallicity by the photometry of RGB stars in the outer halo. They found the metallicity distribution function of the outer halo shows a relatively high metallicity peak at $\rm [M/H] \approx -0.5$, indicating that the stellar metal content in this region is high. This high metallicity is also found in the inner halo and spheroidal regions of M31, which is much higher than the typical metallicity level of $\rm [M/H] \approx -1.5$ in the outer halo of the Galaxy.
Later, \cite{2013ApJ...774....3B} utilized spectroscopic data of planetary nebulae obtained from the 10.4 m Gran Telescopio Canarias, along with 16 outer-disk PNe previously published by \cite{2012ApJ...753...12K}. They discovered that the [O/H] gradient remains unexpectedly flat beyond $R_{\rm GC} \sim$ 20 kpc.

Regarding M33, it is a low-luminosity late-type spiral galaxy. Its proximity ($\sim$800 kpc) and large angular size, combined with a moderate inclination, make it an ideal target for detailed investigation of its metallicity distribution.
Spiral galaxies are widely believed to form in an `inside-out' manner, with the inner regions forming earlier than the outer parts \citep{1976MNRAS.176...31L}. This formation scenario is supported by observations showing a decrease in stellar age with increasing galactocentric radius in some nearby galaxies \citep{2004ApJS..152..175M, 2009ApJ...695L..15W}.
A metallicity gradient naturally arises as a consequence of this inside-out growth.
\cite{1971ApJ...168..327S} was the first to determine radial abundance gradient in M33.
By using near-infrared data from UKIRT/WFCAM, \cite{2008A&A...487..131C} identified a large number of carbon-rich and oxygen-rich AGB stars. With the C-/O-AGB number ratio, it was determined that the metallicity is high in the center, while low in the outer disk and halo. There is an obvious metallicity gradient that decreases linearly with $R_{\rm GC}$, with a spread in [Fe/H] equal to 0.6.
\cite{2009A&A...506.1137C} studied the metallicity gradient of M33 and found that the metallicity of the inner disk up to about 9 kpc decreases linearly by $-0.078 \pm 0.003$ dex kpc$^{-1}$, while the metallicity gradient of the outer disk/halo is relatively gentle, which is related to the physical structure of the spiral arms.
Similarly, \cite{2009ApJ...704.1120U} obtain the average metallicity gradient in the disk as $-0.07 \pm 0.01$ dex kpc$^{-1}$.
\cite{2022ApJ...932...29L} use the low-resolution LAMOST and Keck spectra of blue supergiants in the disk to obtain the metallicity gradient of M33 being $-0.040$ dex kpc$^{-1}$, and the metallicity at the center is 0.11 $\pm$ 0.04 dex.
\cite{2015MNRAS.451.3400B} found that the inner global clusters with $R_{\rm GC}<4.5$ kpc exhibit a steep radial metallicity gradient of $-0.29 \pm 0.11$ dex kpc$^{-1}$, suggesting that these inner global clusters belong to the thick disk.
Lastly, \cite{1986ApJ...305..591M} noted that the M31 halo giant branch is very broad, indicating a large dispersion in metallicity, while the M33 halo giant branch is much narrower and less metal-rich.

In studies of galactic chemical evolution, the stellar age distribution is considered a key factor influencing metallicity gradient. \cite{2015AJ....150..189G} conducted a systematic analysis of the metallicity distribution of red giant branch stars in M31 using optical photometry from the Hubble Space Telescope as part of the PHAT survey. By adopting isochrone interpolation under the assumption of a fixed stellar age, their study revealed the following: (1) metallicity decreases with increasing stellar age, (2) the metallicity gradients remain nearly constant across different assumed ages, and (3) a negative age gradient tends to flatten the metallicity gradient, whereas a positive age gradient results in a steeper one.
Regarding the spatial distribution of stellar ages, many studies have reached broadly consistent conclusions. Based on results of \cite{2015yCat..22150009W}, \cite{2015AJ....150..189G} suggest that the age gradient across the M31 disk is relatively small.
Besides, \cite{2014A&A...570A...6S} analyzed a sample of 62 disk galaxies from the CALIFA survey and found that the age gradient tends to be shallow and negative. Notably, their study also showed that the presence or absence of a bar structure has no significant effect on either gradient.

In summary, previous studies using different data, such as supergiants, red giant branch stars, and Cepheids, provide evidence supporting the existence of metallicity gradients in both M31 and M33 and have estimated some of these gradients.
However, most of the studies focus on limited space.
Here, we aim to investigate the metallicity gradients across as large a region as possible in both M31 and M33 by using the near-infrared color of the TRGB from the comprehensive sample of \cite{2021ApJ...907...18R}.
As introduced in \citet[\citetalias{2024AJ....167..123L} hereafter]{2024AJ....167..123L} which studied the metallicity gradients of the LMC and the SMC, the color of TRGB is an indicator of metallicity.
Besides, TRGB is bright in near-infrared which is insensitive to interstellar extinction.
The application of this method to the LMC and the SMC verified its correctness and has laid a good foundation for studying the metallicity gradients of M31 and M33.

This paper is organized as follows. Section \ref{sec:data} introduces the data, followed by Section \ref{sec:method} to describe the method of detecting the TRGB. The results and discussions are presented in Section \ref{sec:results}, and the summary in Section \ref{sec:summary}.

\section{Data} \label{sec:data}

The near-infrared photometric data is used and taken from the catalog by \cite{2021ApJ...907...18R}. \cite{2021ApJ...907...18R} established a complete and pure sample of evolved red stars in M31 and M33 observed by UKIRT/WFCAM. The observation covered both the disk and halo of M31 and M33, which is larger than the survey by space telescopes.
The outermost extent of M31 reaches up to 170 kpc, while that of M33 exceeds 15 kpc.
Like M31, most previous studies have focused on the metallicity gradients within 100 kpc. However, the catalog by \cite{2021ApJ...907...18R} covers a much broader range, extending the radius of M31 to 170 kpc. As a result, their catalog enables the study of metallicity gradients over a larger radial extent. Additionally, thanks to the continuity of the catalog data, we can investigate metallicity gradients across various radial ranges. This aspect is highlighted in the discussion. Not only have we analyzed the structure within 24 kpc, but we have also obtained a flatter gradient for 24 $<R_{\rm GC}<$ 170 kpc.
They removed the foreground dwarf stars by the $J-H/H-K$ diagram in combination with Gaia astrometric measurements and optical photometry.
The catalog contains 823,189 member stars in M31, and 124,310 in M33, with the limiting magnitude of $K$ $\sim$ 20 mag, which is apparently fainter than the TRGB $K$-band brightness, i.e. 17.62 mag in M31 and 18.11 mag in M33.

\begin{figure}
	\centering
    \includegraphics[scale=0.8]{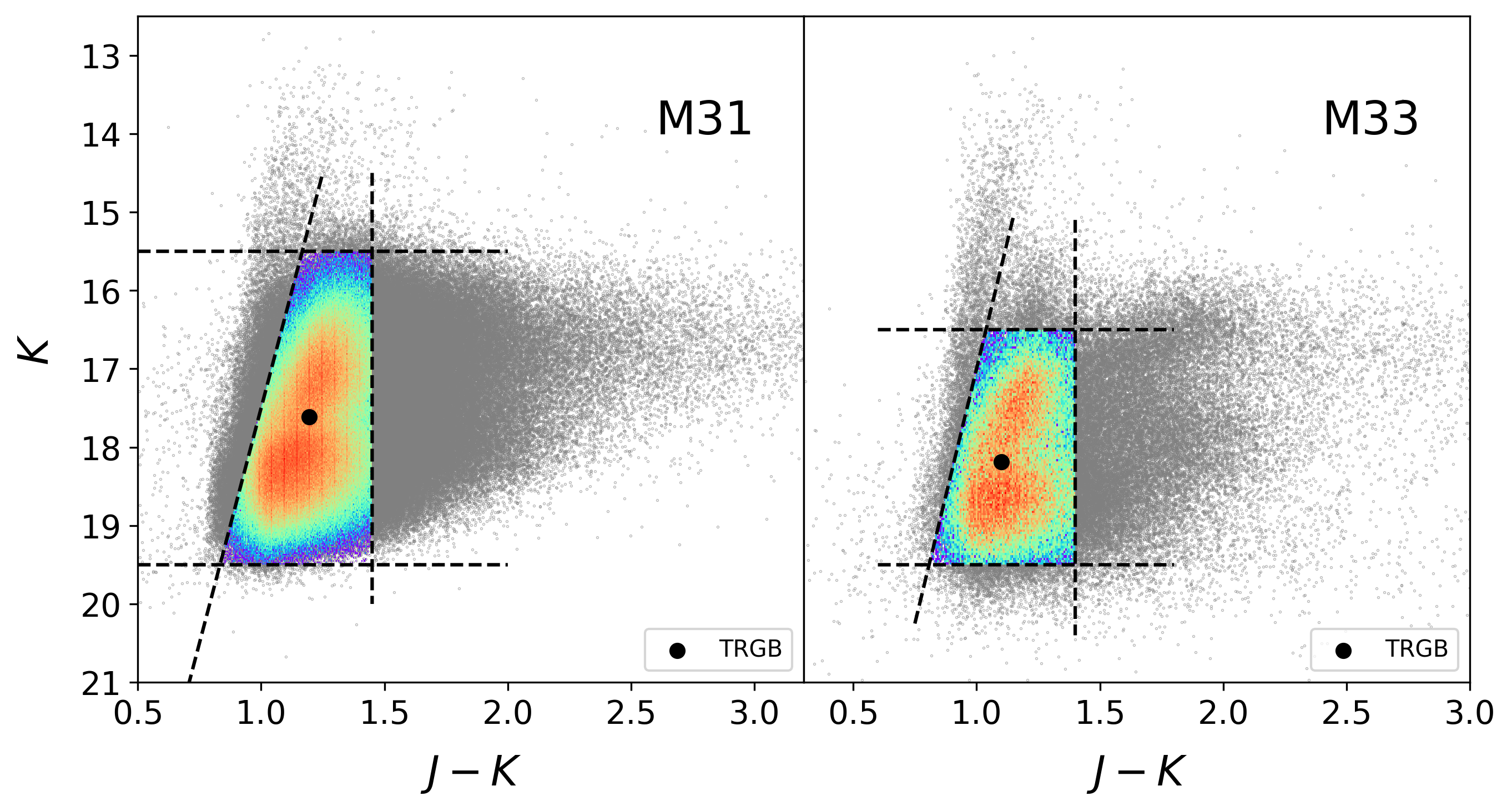}
	\caption{The $J-K$/$K$ diagram of M31 and M33 member stars. The dashed line indicates the area used to determine the color and brightness of TRGB. The black dots indicate the TRGB positions derived in Section \ref{sec: TRGB of M31 and M33}.
    \label{fig:M31_delect}}
\end{figure}

To enhance the accuracy of the TRGB position determination, the boundary condition is applied in the $J-K$/$K$ color-magnitude diagrams (CMDs), illustrated in Figure \ref{fig:M31_delect}, in the same way as performed for the LMC and the SMC in \citetalias{2024AJ....167..123L}.
After applying these constraints, our final sample consists of 678,192 member stars in M31 (Sample-31) and 83,479 member stars in M33 (Sample-33), which is used for later analysis.

\section{Method} \label{sec:method}
To determine the position of TRGB in the $J-K$/$K$ diagram, the Voronoi Binning method \citep{2006MNRAS.368..497D, 2003MNRAS.342..345C} is used to divide the galaxy area into multiple sub-regions, the PN method (Poisson Noise weighted star counts difference in adjacent bins, \cite{2018AJ....156..278G}) is used to find the exact location, and the bootstrap method is used to estimate the uncertainty. The methods are the same as in \citetalias{2024AJ....167..123L} used for LMC and SMC, and the details can be found therein. In this section, we only describe briefly the method and the specific parameters for M31 and M33.

\subsection{The Voronoi Binning} \label{sec:Voronoi Binning}

To study the distribution of metallicity, the entire galaxy area needs to be divided into multiple sub-regions.
Given the highly uneven distribution of stars across the field, using bins with equal spatial intervals is not suitable. Instead, the Voronoi tessellation method considers the uneven distribution and partitions the area into sub-regions with each containing approximately the same number of stars.

Following the spatial distribution of the Sample-31 stars, the boundaries of M31 are defined as:
13$^{\circ}\!\!.$14
$>$R.A.$>$
8$^{\circ}\!\!.$59 and
43$^{\circ}\!\!.$17
$>$Decl.$>$
39$^{\circ}\!\!.$94,
which is divided into 113$\times$80 `pixels' with each pixel of 0$^{\circ}\!\!.$04 by 0$^{\circ}\!\!.$04.
Similarly, the boundaries of M33 are defined by the spatial distribution of the Sample-33 stars as:
24$^{\circ}\!\!.$23
$>$R.A.$>$
22$^{\circ}\!\!.$46 and
31$^{\circ}\!\!.$53
$>$Decl.$>$
29$^{\circ}\!\!.$80,
and divided into 98$\times$96 `pixels' with each pixel of 0$^{\circ}\!\!.$018 by 0$^{\circ}\!\!.$018.
Then the expected number of stars in each bin is set.
On the number of bins and the number of stars in one bin, they anti-correlate. On one hand, the more the bins, the higher spatial resolution the metallicity gradient would have. On the other hand, the more the number of stars in one bin, the more certain the position of TRGB in the color-magnitude diagram. According to \citetalias{2024AJ....167..123L}, the PN method yields reliable result on the TRGB position in the CMD only if the tracers number exceeds 100. Thus, we begin by increasing the star number from 100 to assess the distribution and number of bins in M33. When the star number reaches 300, the region is divided into 263 bins, while it becomes 131 if the target number is 600. When divided into 131 bins, investigating the gradient becomes challenging.
In compromise with the size of each bin and the distribution of bins, the target number for M31 and M33 is set to 300, though the result would not differ significantly when this number is around 200-400.
The expected bins are generated using the adaptive Voronoi binning technique proposed by \cite{2006MNRAS.368..497D}, which is an extension of the Voronoi binning algorithm by \cite{2003MNRAS.342..345C}. To achieve the target of approximately 300 stars per bin, the algorithm starts with a single `pixel' and progressively incorporates adjacent pixels until the bin contains around 300 stars.
As a result, M31 is divided into 1,832 bins, and M33 is divided into 263 bins. The binning results are displayed in Figure \ref{fig: M31_J-K} and Figure \ref{fig: M33_J-K} for M31 and M33 respectively.

\subsection{Detecting the TRGB by the PN method}

The TRGB marks the point at which the star enters the core helium flash, at which point the stellar luminosity suddenly stops increasing and moves to the horizontal branch, reflected in the luminosity function as a breakpoint.
In the CMD, after leaving the main sequence, stars ascend in the RGB, with their luminosity continuously increasing until they reach a maximum at the TRGB. During the helium flash, the stars undergo a rapid, almost `jump-like' transition as they move onto the horizontal branch or settle into the red clump \citep{2005essp.book.....S}. In correspondence with the lifetime in every stage, the star density shows a gradual decrease and subsequent increase from the RGB to the AGB, with the TRGB positioned at the point of lowest density in the $J-K$/$K$ diagram \citep{2002MNRAS.337L..31I, 2021ApJ...907...18R, 2021A&A...649A.110P}.
Early estimates of the TRGB apparent magnitude relied on a basic visual assessment to determine its position in the CMD \citep{1986ApJ...305..591M}.
Several sophisticated techniques have been developed to measure the brightness and colors of the TRGB.
\cite{1993ApJ...417..553L} used an edge detection algorithm to obtain the magnitude of TRGB. They used the Sobel kernel and the magnitude error of the $I$ band was usually 0.1$-$0.2 mag.
\cite{1996ApJ...461..713S} further explored the accuracy and wide application of the Sobel method. They explored the application effect of the Sobel method on different types of data through more detailed tests and error estimation.
\cite{2002AJ....124..213M} used the Maximum Likelihood method to estimate the magnitude of TRGB, but this method requires setting a magnitude range limit for likelihood analysis.
\citet{2018AJ....156..278G} utilized a simple edge detection technique based on the difference in star counts between adjacent bins, weighted by estimated Poisson noise.
\citet{2021ApJ...907...18R} determined the brightness and color of the TRGB in both M31 and M33 by calculating the saddle points.

In this study, we use the Poisson Noise weighted star counts difference in adjacent bins, as suggested by \citet{2018AJ....156..278G}, to identify the TRGB in both M31 and M33, which suits our data as demonstrated in \citetalias{2024AJ....167..123L}.

\subsection{The Bootstrap error analysis}

The uncertainty in the CMD is estimated by using the bootstrap method. Introduced by \citet{Efron1979}, the bootstrap resampling technique involves generating numerous data sets by randomly drawing data points with replacement from the original sample, each set containing the same number of points as the original. Some points may be duplicated, while others may be omitted.

In this study, we use a no-relaxation sampling method (parameter: replace=0), where 80\% of the total sample is adopted for each resampling iteration to apply the PN method for calculating the color index and magnitude of the TRGB. This procedure is repeated 2000 times, resulting in 2000 sets of TRGB color index and magnitude values.

From these 2000 sets of results, the expected value $\mu_{J-K}$ and its standard deviation $\sigma_{J-K}$ are divided by fitting a Gaussian function to the distribution of $J-K$. We then select the results within one standard deviation $\sigma_{J-K}$ of the mean to fit $K$, which yields the expected value $\mu_{K}$ and its standard deviation $\sigma_{K}$. This method ensures robustness and reliability in estimating the key parameters of interest.

\section{Result and Discussion} \label{sec:results}

\subsection{TRGB of M31 and M33} \label{sec: TRGB of M31 and M33}

The method is first verified in the entire galaxy by selecting 80\% of the data using a no-replacement sampling method to perform the PN algorithm, subsequently obtaining a set of $J-K$ and $K$ values. 
This process is repeated 2000 times, yielding 2000 sets of $J-K$ and $K$. We adopt the median and standard deviation as the statistical measures for $J-K$ and $K$. The derived TRGB position for M31 is $J-K$ = 1.195 $\pm$ 0.002 and $K$ = 17.615 $\pm$ 0.007, while for M33, $J-K$ = 1.100 $\pm$ 0.003 and $K$ = 18.185 $\pm$ 0.053.
The TRGB positions are marked with black dots in Figure \ref{fig:M31_delect}.
In comparison, \citet{2021ApJ...907...18R} used the saddle points method to determine the TRGB position in the CMD with the same data sample, reporting $J-K = 1.20$ and $K = 17.62$ for M31, and $J-K = 1.09$ and $K = 18.11$ for M33.
The TRGB positions determined in this study show a high consistency with these results, proving the reliability of our approach.

\subsection{Distribution of J-K}

As mentioned in Section \ref{sec:method}, the Voronoi method is used to divide M31 into 1832 bins and M33 into 263 bins, and then, the PN and bootstrap methods are combined to obtain the position and its uncertainty of TRGB in each bin.
The spatial distribution of resultant $J-K$ in M31 and M33 is shown in Figure \ref{fig: M31_J-K} and Figure \ref{fig: M33_J-K}, respectively.
A radial gradient is evident in M31, whereas the trend in M33 is comparatively weaker. This variation in $J-K$ may reflect a decrease in metallicity from the inner to the outer region of the galaxies.

\subsubsection{J-K distribution in M31}

\begin{figure}
    \centering
    \includegraphics[scale=0.65]{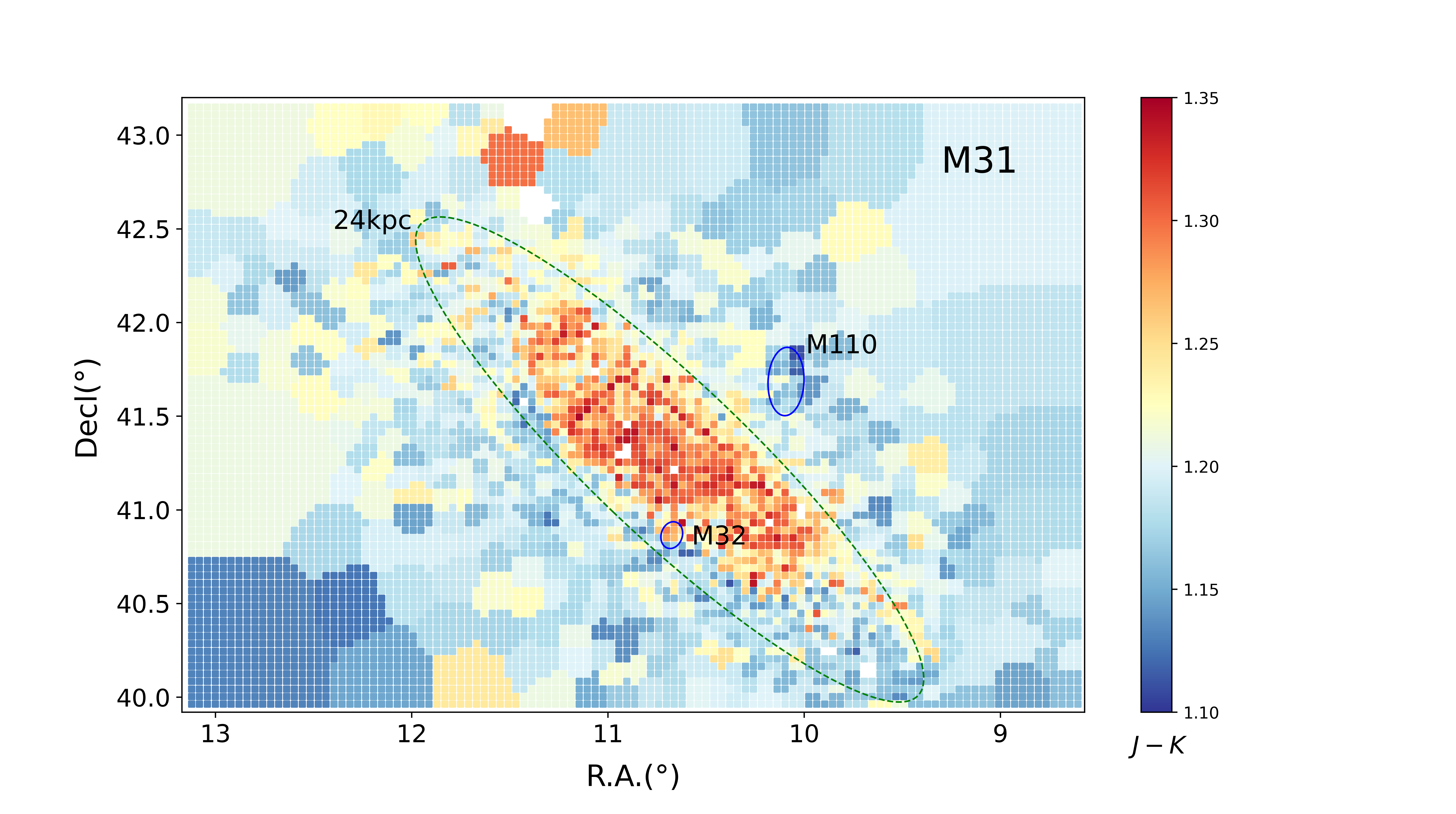}
    \caption{The $J-K$ distribution for M31. The smallest squares represent the initial pixel unit and the irregular polygon including multiple pixel units of the same color denotes one final bin with about 300 stars by the Voronoi binning methods. In total, there are 1832 bins.}
    \label{fig: M31_J-K}
\end{figure}

Globally, the $J-K$ value distribution of M31 is high in the center and low close to the edge, with a clear negative gradient from the inner to the outer region of the galaxy. This phenomenon is consistent with previous conclusions as described in Section \ref{sec:introduction}.
In the distribution of $J-K$, it can be seen that the values in the south are slightly larger than those in the north, and slightly higher in the western than the eastern outer disk.
Specifically, from R.A.$\sim$9$^{\circ}\!\!.$5 in the west, $J-K$ is $\sim$1.2, while to R.A.$\sim$12$^{\circ}$ in the east, $J-K$ is $\sim$1.15.
This is consistent with the study of the star formation history of M31 by \cite{2003AJ....126.1312W}, which may be related to the interaction with NGC 205 to the north-west and M32 to the south of the galaxy disk.
There are some locations where $J-K$ values are similarly clustered in Figure \ref{fig: M31_J-K}, which may be the structural features of the galaxy M31, very likely the dust rings.

\cite{2014ApJ...780..172D} identified three dust rings at 5.6 kpc, 11.2 kpc, and 15.1 kpc respectively by using the data from the Spitzer Space Telescope and Herschel Space Observatory, in combination with a physical dust model.
\cite{2014ApJ...786..130L} confirm that the dust rings are at 5 kpc, 12 kpc, and 15 kpc, meanwhile \cite{2011A&A...536A..52M} instead suggest the rings at 0.7 kpc and 10 kpc.

\begin{figure}
    \centering
    \includegraphics[scale=0.9]{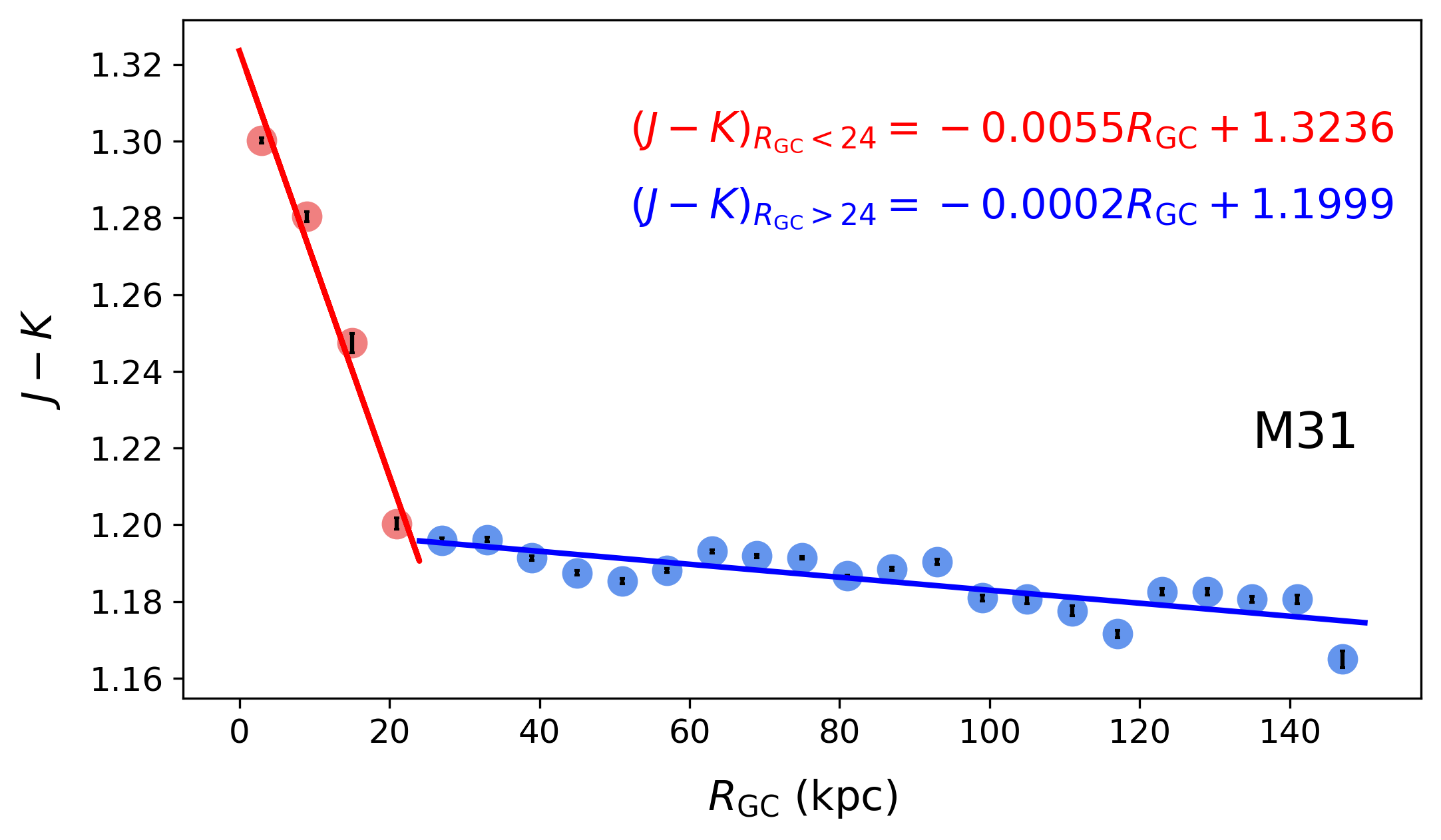}
    \caption{The variation of the $J-K$ with the $R_{\rm GC}$ from the galactic center in M31. The red line is fitted to the red points, and the blue line to the blue points.}  
    \label{fig: M31_Dis_J-K}
\end{figure}

To study the metallicity gradient within the galaxy, the coordinates are converted to the galactocentric frame.
The conversion takes into account both the spherical coordinates and the de-projection by following \cite{2009A&A...506.1137C} and \citetalias{2024AJ....167..123L}.
The adopted parameters are the following: the galaxy center at (10$^{\circ}\!\!.$684793, 41$^{\circ}\!\!.$269065), position angle = 45\degr \citep{2003AJ....125..525J}, inclination = 77$^{\circ}\!\!.$5 \citep{2001ChPhL..18.1420M}, and distance = 770 kpc \citep{1990ApJ...365..186F}.
Given the small uncertainty in the distance to M31 and the fact that TRGB metallicity primarily depends on color rather than luminosity, the distance uncertainty is expected to have a negligible impact on the analysis presented in this study.
Afterward, the radius is measured from the center in the galactocentric frame. Starting from the center, an annulus is taken every 6 kpc and the median of $J-K$ in this annulus is used as the representative at this radial distance. The results are displayed in Figure \ref{fig: M31_Dis_J-K} which clearly shows two different trends. When the distance is not larger than 24 kpc, $J-K$ shows a clearly decreasing trend, which can be fitted by $J-K = - 0.0055R_{\rm GC} + 1.3236$. When the distance is greater than 24 kpc, $J-K$ is relatively flat overall, extending to about 150 kpc at its farthest point which can be fitted by $J-K = - 0.0002R_{\rm GC} +1.1999$. The black bar represents the dispersion. 
The position of $R_{\rm GC}=24$ kpc is denoted by the dotted circle in Figure \ref{fig: M31_J-K}.

\begin{figure}
    \centering
    \includegraphics[width=0.8\linewidth]{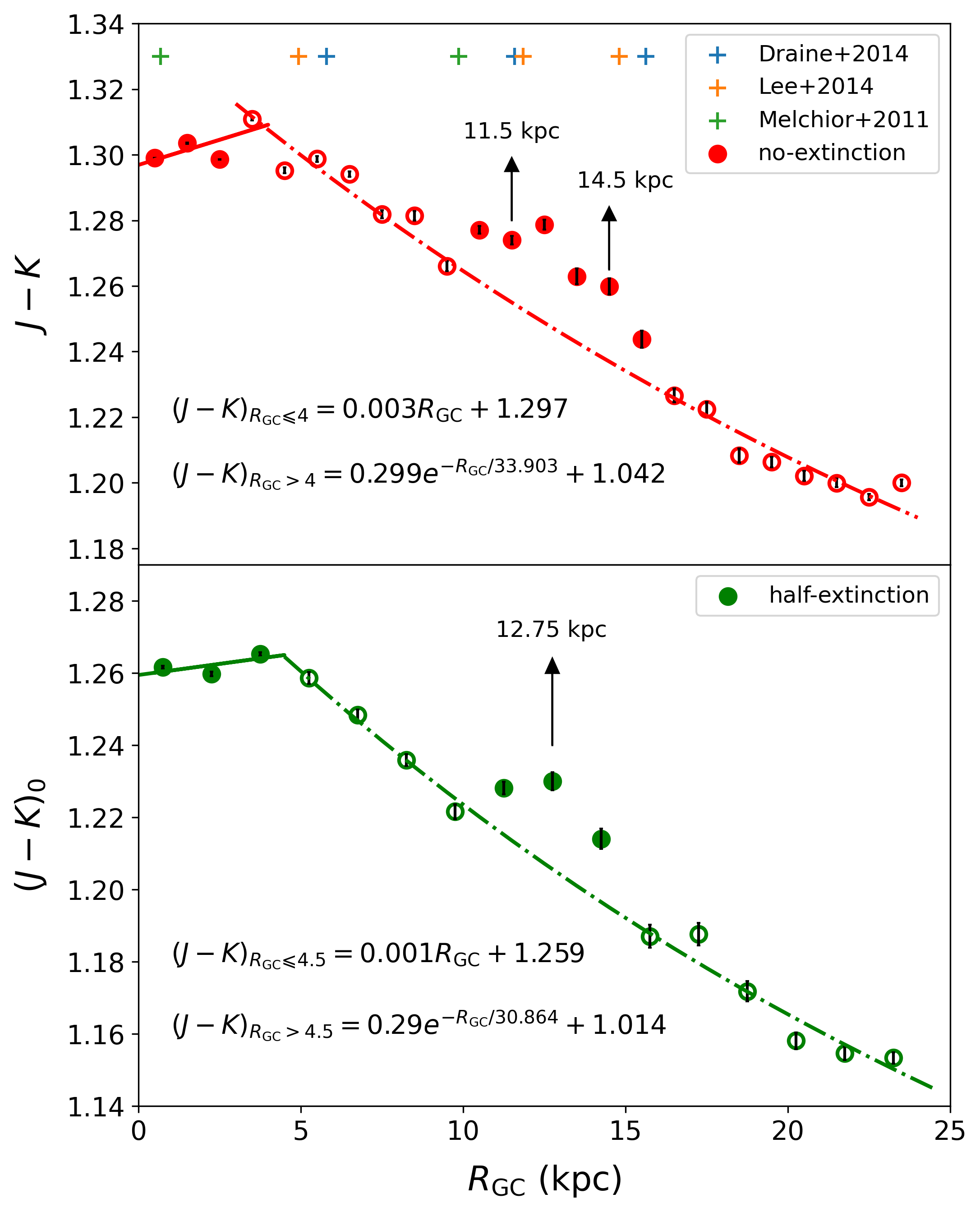}
    \caption{Distribution of $J-K$ and galactocentric radius of 0$-$24 kpc in M31. The top panel is for no-extinction data. The first four points are fitted to get the red straight line, and the red dotted line is fitted by the open red circle. The black bar represents the dispersion. Here are two bumps at 11.5 kpc and 14.5 kpc. The color plus sign is the dust ring position given in different works. The bottom panel is for half-extinction data.}
    \label{fig: M31_25dis_J-K}
\end{figure}

A more detailed analysis is performed for the regions with a galactocentric distance within 24 kpc.
With the step size of 1 kpc, the $J-K$ gradient is shown in Figure \ref{fig: M31_25dis_J-K} where the black bar denotes the dispersion.
In 0$-$4 kpc, $J-K$ shows an increasing trend. A linear fit to the first four points yields a gradient of 0.003 dex kpc$^{-1}$.
Within this range, $J-K$ increases from 1.297 at $R_{\rm GC} \sim 0$ to 1.309 at $R_{\rm GC} \sim 4$ kpc, i.e. a rise of 0.012 mag, roughly 10\% of the total $J-K$ variation ($\sim$0.11 mag).
This is consistent with the shallow positive metallicity gradient in the inner part of M31 given by \cite{2018A&A...618A.156S} and \cite{2023ApJ...952...23G}.
From 4 to 24 kpc, $J-K$ decreases with $R_{\rm GC}$.
Here we can clearly see two bumps at 11.5 kpc and 14.5 kpc respectively. These bumps can be the locations of the dust rings/spiral arms. Neglecting the points of these two structures, the trend is fitted by exponential function by using the red open circles which yields $J-K$ = 0.299 $\times$ e$^{-R_{\rm GC}/33.903}$ + 1.042.

\begin{figure}
    \centering
    \includegraphics[width=0.7\linewidth]{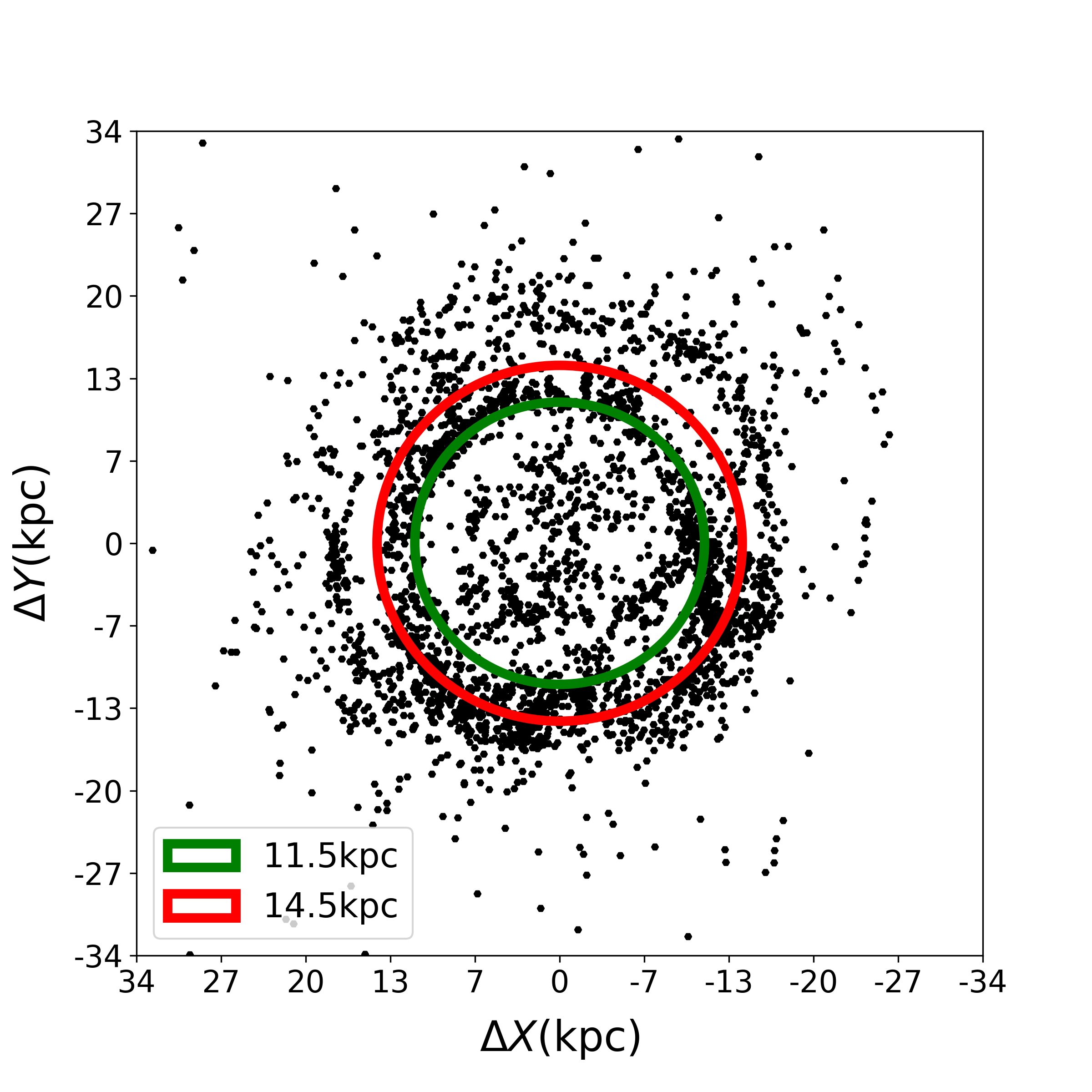}
    \caption{Distribution of dust rings in M31. Red and green represent the position of 14.5 kpc dust ring and 11.5 kpc dust ring. The black dots are the distribution of Sample-R-M31.}
    \label{fig: M31_RSG_ring}
\end{figure}

The locations of the structures are compared with previous works.
The plus signs of different colors at the top of Figure \ref{fig: M31_25dis_J-K} represent the dust rings given by \cite{2014ApJ...780..172D}, \cite{2014ApJ...786..130L}, and \cite{2011A&A...536A..52M} (after distance conversion).
The structures at 11.5 kpc and 14.5 kpc are at a very similar radius to that of \cite{2014ApJ...786..130L} and \cite{2014ApJ...780..172D}.
Spiral arms are generally considered as prominent star-forming regions in galaxies. Similarly, dust rings, which accumulate substantial amount of interstellar dust, also serve as active sites of star formation. These two structural components sometimes exhibit spatial overlap in their distribution within M31 \citep{2014ApJ...786..130L, 2012ApJ...756...40S}.
In terms of spatial structure, the RSGs of M31 and M33 very much follow the spiral arms \citep{2021AJ....161...79M, 2021ApJ...907...18R}. Their locations coincide highly with the spiral arms, which are expected for RSGs as young massive stars. Therefore, a sample of red supergiants is chosen from \cite{2021ApJ...907...18R} to test whether our structure is consistent with spiral arms/dust rings.
The criteria are the stars tagged as `rsg', $G$mag in Gaia greater than 14 mag, parallax $<$ 0.25, or parallax $>$ 0.25 and error/parallax $>$ 0.2, or there is no parallax value.
A total of 3295 sources (hereinafter referred to as Sample-R-M31) are obtained, which are denoted by black solid points in Figure \ref{fig: M31_RSG_ring} (after de-projection).
The structures in Figure \ref{fig: M31_25dis_J-K} are marked in Figure \ref{fig: M31_RSG_ring}, where the black dots are from Sample-R-M31. The ring structure can be seen intuitively from the density of the black dots. The two structures of 11.5 kpc and 14.5 kpc that we obtained are more consistent with the distribution of red supergiants.

\subsubsection{J-K distribution in M33}

\begin{figure}
    \centering
    \includegraphics[scale=0.8]{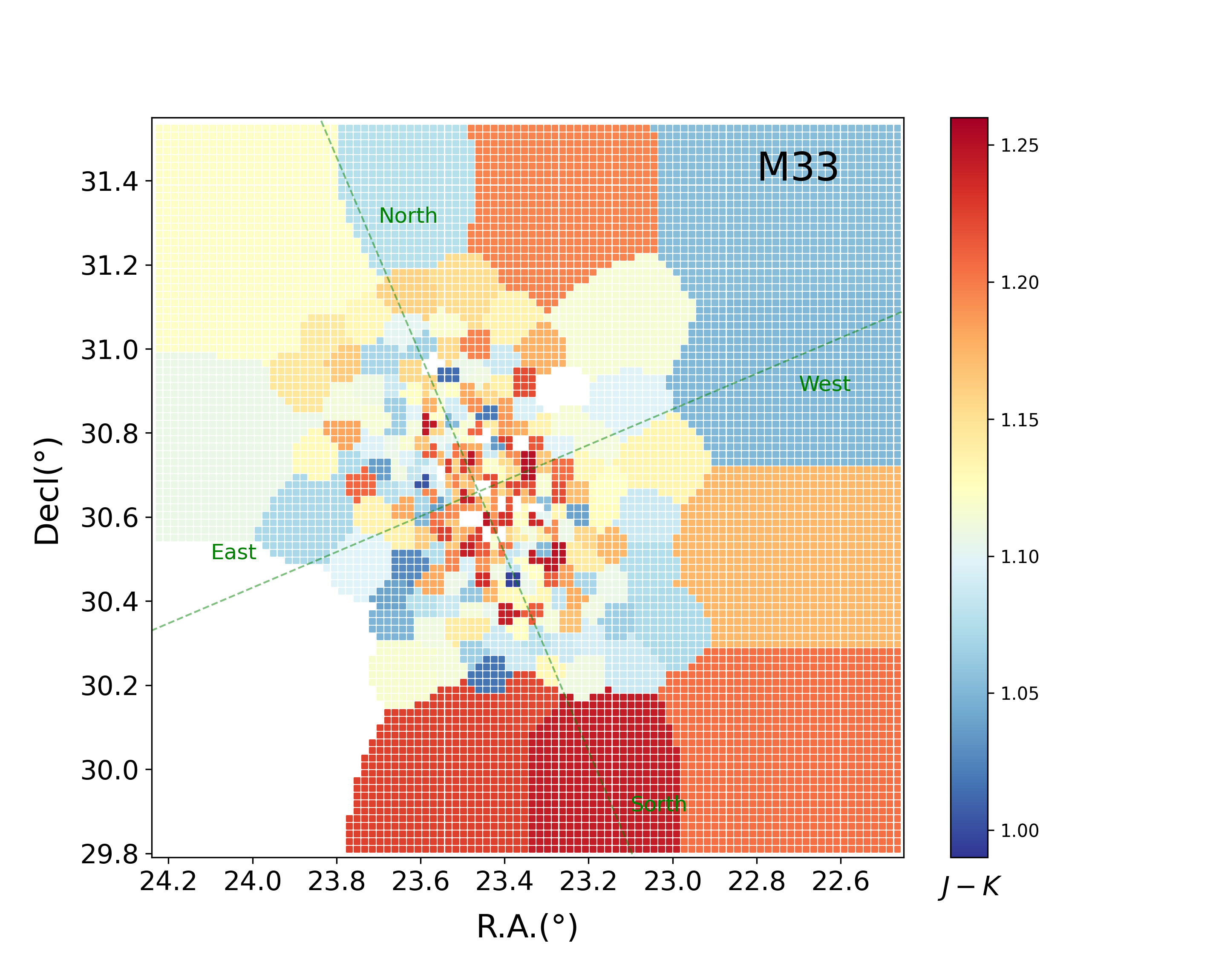}
    \caption{The same as Figure \ref{fig: M31_J-K}, but for M33 and there are 263 bins.}
    \label{fig: M33_J-K}
\end{figure}

For M33, the coordinates are also converted to galactocentric frame by taking into account the spherical coordinates and the de-projection as in \cite{2009A&A...506.1137C} and \citetalias{2024AJ....167..123L}.
The adopted parameters are the following: the galaxy center is (23$^{\circ}\!\!.$462042, 30$^{\circ}\!\!.$660222), position angle = 23\degr \citep{1991rc3..book.....D}, inclination = 53\degr \citep{2009ApJ...696..729M}, and distance = 794 kpc \citep{2004MNRAS.350..243M}.
The $J-K$ distribution of M33 is shown in Figure \ref{fig: M33_J-K}, which is larger in the center and decreasing gradually outward.
In addition, there is a gathering of similar $J-K$ values at some locations. This should be related to the structure of the spiral arms.

\begin{figure}
    \centering
    \includegraphics[scale=0.75]{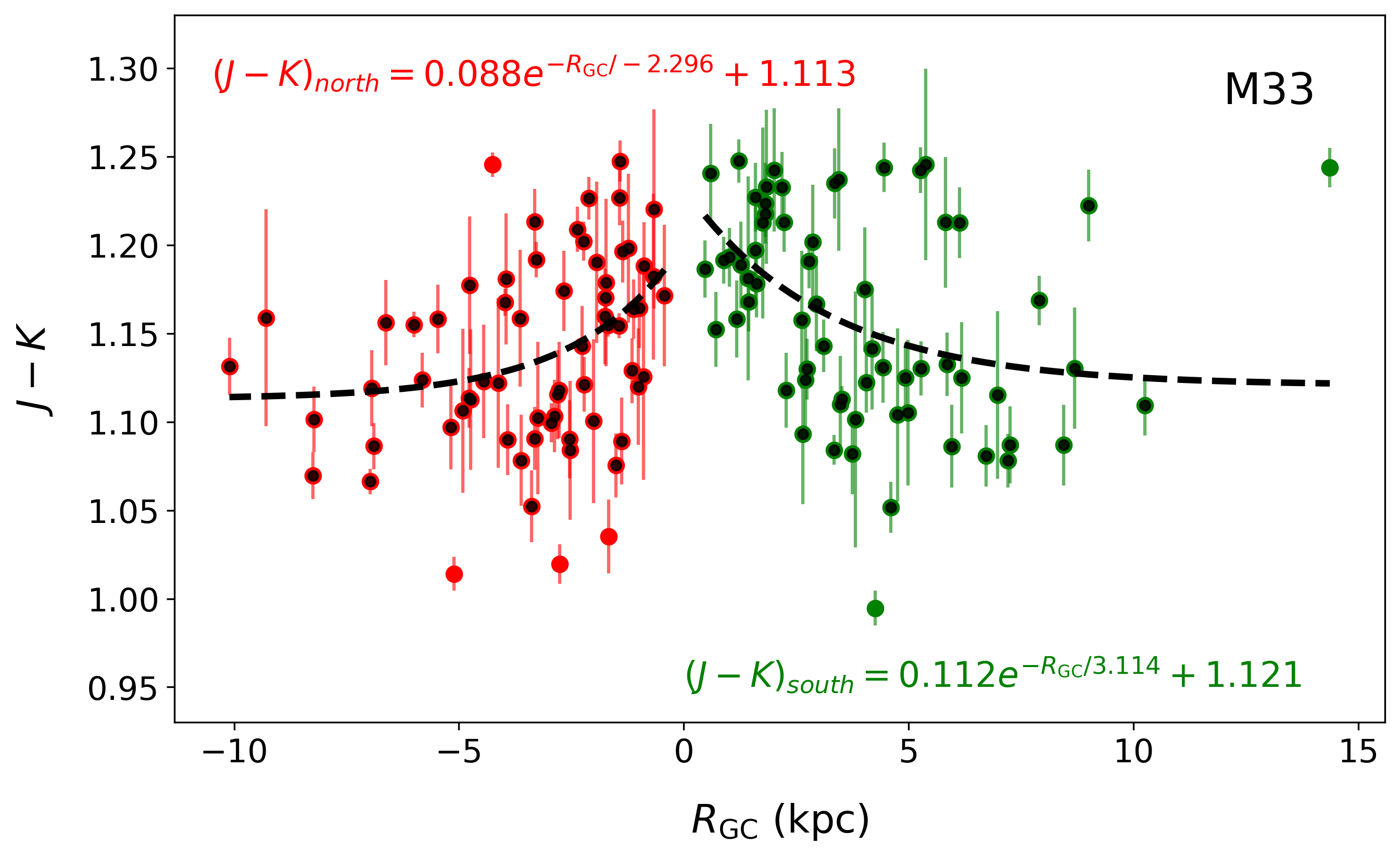}
    \caption{$J-K$ gradient for north and south in M33. Red represents the north, and green represents the south. Black dots are within $2\sigma$.}
    \label{fig: M33_PA_67_J-K_dis}
\end{figure}

\begin{figure}
    \centering
    \includegraphics[scale=0.76]{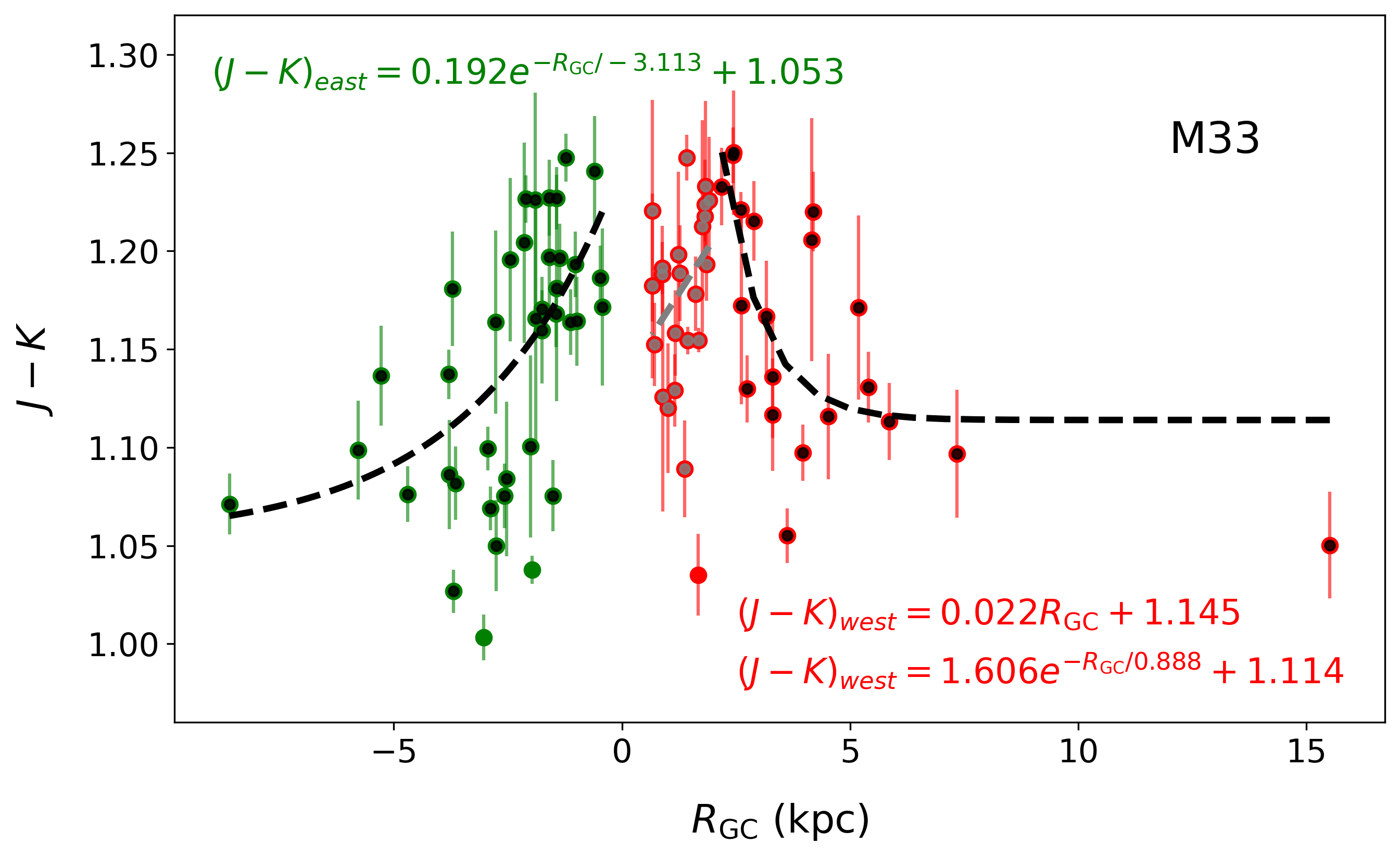}
    \caption{$J-K$ gradient for east and west in M33. Green represents the east and red represents the west.}
    \label{fig: M33_PA67+90_J-K_dis}
\end{figure}

We divide M33 into four quadrants along the position angle direction and its perpendicular direction, using the galaxy center as the origin.
Along the direction of the position angle is the north and south, perpendicular to the position angle is the east and west (see Figure \ref{fig: M33_J-K}).
The bins within 0$^{\circ}\!\!.$1 from the central-line in the north-south or east-west direction are selected to study the radial variation.
Figure \ref{fig: M33_PA_67_J-K_dis} shows the results for the north and south. Taking the north direction as an example, the red points are pre-fitted with an exponential function, and then the points within 2$\sigma$ (black points) are selected for re-fitting to obtain the final function: $J-K=0.088 \times e^{-R_{\rm GC}/-2.296}+1.113$. Similarly, the fitting function for the south direction (green points) is: $J-K=0.112 \times e^{-R_{\rm GC}/3.114}+1.121$.
The results of east and west are displayed in Figure \ref{fig: M33_PA67+90_J-K_dis}. The fitting function of green points is $J-K=0.192 \times e^{-R_{\rm GC}/-3.113}+1.053$.
The red points clearly show two trends, so we make a piecewise function. The function for 0$-$2 kpc is monotonically increasing with a gradient of 0.022 mag kpc$^{-1}$. For $R_{\rm GC} >$ 2 kpc, it is an exponential function: $J-K=1.606 \times e^{-R_{\rm GC}/0.888}+1.114$.
In general, $J-K$ in the east, south, and north directions all decrease with $R_{\rm GC}$, but the magnitudes vary.
The decreasing trend of $J-K$ from the center to the periphery of M33 supports the `inside-out' formation mechanism of M33. In the west, which is the direction toward M31, $J-K$ first increases and then decreases. This could be related to the disturbance between M33 and M31. This interaction could affect the gas distribution, which in turn influences stellar formation and the metallicity distribution \citep{2008A&A...487..161G}.

\subsection{Metallicity Gradient}

\begin{figure}
    \centering
    \includegraphics[scale=0.75]{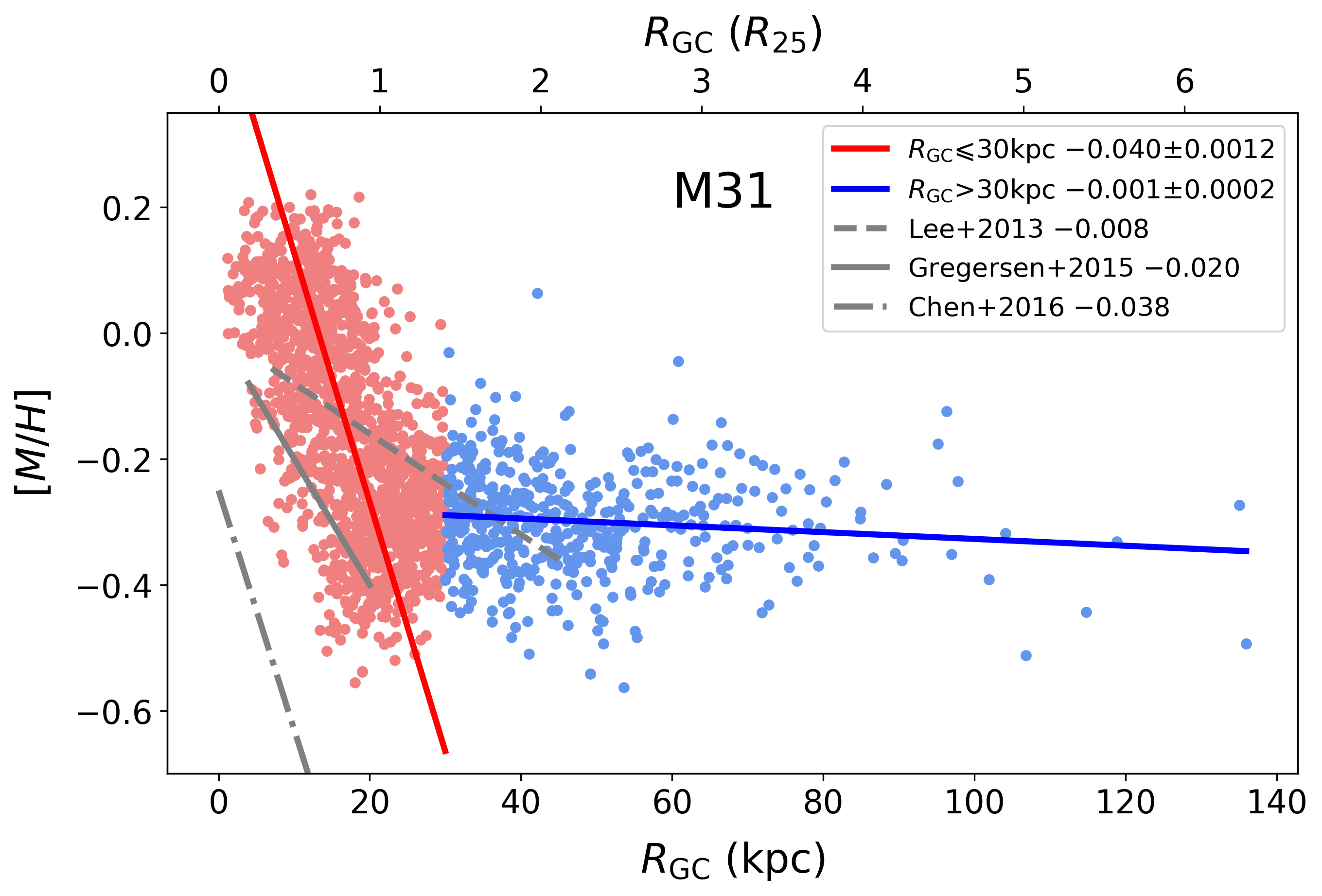}
    \caption{The variation of metallicity with $R_{\rm GC }$ in M31. Red dots are the bins within 30 kpc, and blue dots are the bins greater than 30 kpc. The dashed gray line is $-$0.008 dex kpc$^{-1}$ derived in the $R_{\rm GC}$ range of 7$-$45 kpc given by \cite{2013ApJ...777...35L}, the solid gray line is $-$0.020 dex kpc$^{-1}$ derived in the $R_{\rm GC}$ range of 4$-$20 kpc by \cite{2015AJ....150..189G}, and the dotted gray line is $-$0.038 dex kpc$^{-1}$ derived in the $R_{\rm GC}$ range of 0$-$30 kpc by \cite{2016AJ....152...45C}.}
    \label{fig: M31_Z_dis}
\end{figure}

Both $J-K$ and $K$ serve as metallicity indicators.
$K$ is related to both metallicity and age. \cite{2015AJ....150..189G} shows that while the mass-weighted average age of the M31 disk is relatively old, typically around 7$-$10 Gyr, the average age of stars along the upper RGB is normally younger than the mass-weighted average stellar age. They calculate the mean age of RGB stars to be $\sim$4 Gyr throughout the disk, and the age of TRGB should be comparable.
By using the relationship between age and $M_{K}$ in \cite{2005MNRAS.357..669S} for [Fe/H] = 0.0 dex, the increase in $M_{K}$ is about 0.15 mag if age changes from 3 Gyr to 5 Gyr. On the other hand, $M_{K}$ would vary by about 0.3 mag if the metallicity changes from $-$0.38 to $-$0.68. It can be seen that the $K$ magnitude depends mainly on metallicity but also on the age. As mentioned in \citetalias{2024AJ....167..123L}, the apparent $K$ magnitude is influenced by the vertical depth in the galaxy to which extent is unknown. Consequently, the $K$ brightness is not used to study the metallicity gradient.
Furthermore, \citet{2020ApJ...891...57F} highlight that most of the color width of old ($>$4 Gyr) RGB stars result from a spread in metallicity other than in age.
But , the TRGB brightness in the $K$-band can be influenced by vertical depth, whereas $J-K$ is less sensitive to spatial variations. So the spatial structure of M31 and M33 can smooth out trends in the $K$-band, making $K$ a less reliable metallicity tracer.
The color index $J-K$ of the TRGB serves as a reliable metallicity indicator, in the way that it increases with metallicity.
An additional advantage of using $J-K$ as a metallicity indicator is its little sensitivity to reddening \citep{2000AJ....119..727B}.
When observed at the near-infrared wavelengths at which a TRGB star's spectral energy distribution peaks, the TRGB brightness is strongly (and linearly) correlated with the metallicities/colors of coeval TRGB stars \citep{2024ApJ...975..111H}.
The conversion from $J-K$ to [M/H] has been investigated by a few works which are generally in agreement with each other \citep{2004A&A...424..199B, 2004MNRAS.354..815V, 2025ApJ...980..218S}. 
Here we adopt the relationship $J-K=0.3 \times [M/H]+1.28$ suggested by \cite{2004MNRAS.354..815V}, the same as in \citetalias{2024AJ....167..123L}.
The results of the variation of metallicity with galactocentric radius in M31 and M33 are shown in Figure \ref{fig: M31_Z_dis} and Figure \ref{fig: M33_Z_dis}, respectively.
Since the conversion is linear, the structure is fully the same as that of the color index $J-K$ discussed in the previous subsection, but the amplitude differs.

\subsubsection{M31}

The metallicity gradient is calculated by using the least square method with a segmented linear function.
The slope of metallicity with $R_{\rm GC}$ is $-0.040 \pm 0.0012$ dex kpc$^{-1}$ for $R_{\rm GC}$ $<$ 30 kpc and $-0.001 \pm 0.0002$ dex kpc$^{-1}$ for $R_{\rm GC}$ $>$ 30 kpc. The range extends to 140 kpc.
The Pearson correlation coefficient is $-$0.63 for $R_{\rm GC}<30$ kpc, and $-$0.11 for $R_{\rm GC}>30$ kpc. 
Within 30 kpc, an evident correlation is observed between $R_{\rm GC}$ and $[M/H]$.
For comparison, the dashed gray line in Figure \ref{fig: M31_Z_dis} represents $-$0.008 dex kpc$^{-1}$ derived in the $R_{\rm GC}$ range of 7$-$45 kpc by \cite{2013ApJ...777...35L}, the solid gray line is $-$0.020 dex kpc$^{-1}$ derived in the $R_{\rm GC}$ range of 4$-$20 kpc by \cite{2015AJ....150..189G}, and the dotted gray line is $-$0.038 dex kpc$^{-1}$ derived in the $R_{\rm GC}$ range of 0$-$30 kpc by \cite{2016AJ....152...45C}.
Their metallicity ranges differ, possibly due to variations in the tracer populations used or differences in the treatment of interstellar extinction.
Our slope is slightly steeper than theirs.
The central regions of galaxies are under strong gravitational constraints and weak centrifugal forces, leading to the concentration of gas and dust. This facilitates accelerated star formation to enhance the metallicity. In the outer regions, where the gravitational force is weaker, some gas may be expelled by centrifugal forces, resulting in a more diffuse distribution of gas and dust.
As a result, the lower star formation efficiency in the outer regions of the galaxy leads to low metallicity.
The negative gradient can be understood theoretically.
\cite{2010A&A...520A..35M} modeled the distribution of star formation rates in the Milky Way, M31, and M33, and their results align with observational data from \cite{1991ARA&A..29..129R} and \cite{2000ApJ...541..597H}. These studies indicate that the central regions of galaxies have higher star formation rates and consequently higher metal abundances, while the outer regions have lower star formation rates and lower metallicities.
In addition, based on long-slit spectroscopy and integral field unit spectroscopy, most spiral galaxies in the local Universe exhibit negative metallicity gradients in gas and stars within their optical radius, that is, the centre of a galaxy has a higher metallicity than the outskirts \citep{2023A&A...679A..83K, 1994ApJ...420...87Z, 2019MNRAS.488.3826B, 2022ApJ...940...32B}.
While star formation rates are crucial in driving chemical evolution, they alone are insufficient to fully explain the observed metallicity distributions. Time scales play a vital role in this context.
In the inside-out formation model, the central regions, which formed earlier, experienced a higher rate of stellar processing and metal enrichment, leading to a higher metallicity in the nucleus. In contrast, the outer disk, which formed later, had less time to accumulate metals, resulting in a lower metallicity \citep{2021MNRAS.502.5935S}.

\subsubsection{M33} \label{sec: M33_Z}

\begin{figure}
    \centering
    \includegraphics[scale=0.8]{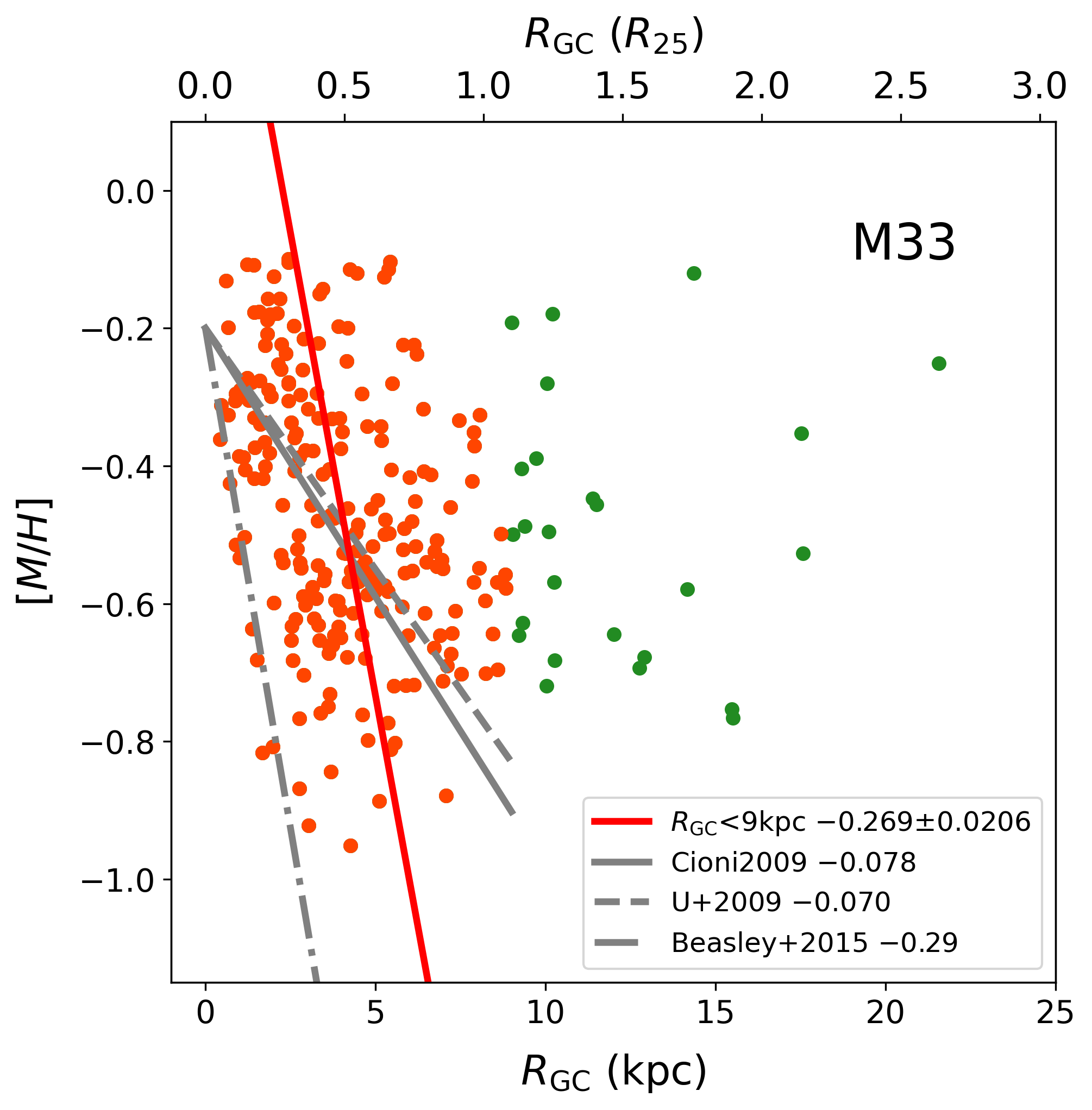}
    \caption{The variation of metallicity with $R_{\rm GC }$ in M33. Red dots are the bins within 9 kpc, and green dots are the bins greater than 9 kpc. The solid gray line is $-$0.078 dex kpc$^{-1}$ derived in the $R_{\rm GC}$ range of 0$-$9 kpc by \cite{2009A&A...506.1137C}, the dashed grey line is $-$0.070 dex kpc$^{-1}$ derived in the same $R_{\rm GC}$ range by \cite{2009ApJ...704.1120U}, and the dotted gray line is $-$0.29 dex kpc$^{-1}$ derived within $R_{\rm GC}<4.5$ kpc by \cite{2015MNRAS.451.3400B}.}
    \label{fig: M33_Z_dis}
\end{figure}

The variation of metallicity with galactocentric radius is shown in Figure \ref{fig: M33_Z_dis}. With reference to previous works, the range of 0$-$9 kpc is selected to calculate the metallicity gradient.
The result is $-$0.269 $\pm$ 0.0206 dex kpc$^{-1}$.
The Pearson correlation coefficient for the data points between 0 and 9 kpc was calculated to be $-$0.35, indicating a moderate negative correlation between $R_{\rm GC}$ and $[M/H]$.
In contrast, the metallicity beyond 9 kpc shows significant scatter and appears relatively flat.
The solid gray line in Figure \ref{fig: M33_Z_dis} is $-$0.078 dex kpc$^{-1}$ derived in the $R_{\rm GC}$ range of 0$-$9 kpc by \cite{2009A&A...506.1137C}, the dashed gray line is $-$0.070 dex kpc$^{-1}$ derived in the same $R_{\rm GC}$ range by \cite{2009ApJ...704.1120U},
and the dotted gray line is $-$0.29 dex kpc$^{-1}$ derived within $R_{\rm GC}<4.5$ kpc by \cite{2015MNRAS.451.3400B}.
Our gradient is larger than some previous works and is in good agreement with the result of \cite{2015MNRAS.451.3400B}.
The steeper metallicity gradient in the inner disk and the flatter gradient in the far region of M33 may be related to the physical structure of the spiral arms. \cite{2007A&A...470..843M} showed that the metallicity gradient of M33 has been slowing down over the past 8 Gyr.

Compared to M31, the metallicity gradient of M33 appears to be much steeper \citep{1991AJ....102..927A, 1982ApJ...254...50B}.
This phenomenon may be related to the mass and luminosity differences between M31 and M33.
M31 is a large spiral galaxy with significantly higher mass and luminosity compared to M33.
\cite{2015MNRAS.448.2030H} studied the metallicity gradients in 49 local field star-forming galaxies and noted that when the metallicity gradients are expressed in dex kpc$^{-1}$, galaxies with lower mass and luminosity tend to have steeper gradients. However, when the metallicity gradients are expressed in dex $R^{-1}_{25}$, no correlation is found between metallicity gradients and stellar mass or luminosity.
The dependence of metallicity gradient on galactic mass while measuring the metallicity gradient in absolute scale (kpc) or relative scale ($R_{\rm 25}$) can be understood as a size effect. If galaxies with steeper metallicity gradient in dex kpc$^{-1}$ are smaller in their physical sizes (small $R_{\rm 25}$), then the steep dex kpc$^{-1}$ metallicity gradients would be compensated when the galaxy sizes are taken into account.
In their work, low-luminosity galaxies have smaller $R_{\rm 25}$ than high-luminosity galaxies, indicating that the steep dex kpc$^{-1}$ metallicity gradients in low-mass galaxies could be associated with their small physical sizes. These results imply that the evolution of metallicity gradients is closely related to the growth of galaxy size.
In Figures \ref{fig: M31_Z_dis} and \ref{fig: M33_Z_dis}, we also define the radius using $R_{\rm 25}$. For M31, $R_{\rm 25}$ = $95^{\prime}\!.30 = 21.3$ kpc, and for M33, $R_{\rm 25}$ = $35^{\prime}\!.40 = 8.18$ kpc \citep{1991rc3..book.....D, 1990ApJ...365..186F, 2004MNRAS.350..243M}.
However, since the correlation between $R_{\rm GC}$ and $[M/H]$ in M33 with the Pearson coefficient of $-$0.35 is much weaker than that in M31 with the Pearson coefficient of $-$0.63, it cannot be simply concluded that the metallicity gradient in M33 is steeper than in M31.

\subsubsection{Comparison with the LMC and the SMC} \label{sec: compare}

The LMC and SMC are the most massive and closest dwarf satellite galaxies of the Milky Way, located at distances of approximately 50 kpc and 61 kpc, respectively \citep{2014AJ....147..122D, 2015AJ....149..179D}. M31 and M33 are nearby spiral galaxies in the Local Group and serve as important analogs to the Milky Way in studies of galaxy evolution. All four galaxies benefit from extensive, high-quality photometric datasets, enabling detailed investigations of their stellar populations and chemical properties. In particular, the catalogs published by \cite{2021ApJ...907...18R, 2021ApJ...923..232R} provide uniformly processed and well-calibrated near-infrared photometry for stars across these galaxies, offering ideal conditions for robust measurements of their metallicity gradients.
This section compares the metallicity gradients of the LMC, SMC, M31, and M33, with the aim of exploring the similarities and differences in metallicity gradients between irregular and spiral galaxies on a localized scale.

In \citetalias{2024AJ....167..123L}, the metallicity gradients of the LMC and the SMC are derived.
Both galaxies are divided into four distinct directions with the metallicity gradient ranging from $-$0.006 dex kpc$^{-1}$ to $-$0.010 dex kpc$^{-1}$ for LMC, and from $-$0.003 dex kpc$^{-1}$ to $-$0.017 dex kpc$^{-1}$ for SMC. Compared to the LMC and the SMC, the metallicity gradient is much more pronounced in M31 ($-$0.040 dex kpc$^{-1}$) and in M33 ($-$0.269 dex kpc$^{-1}$).
There may be several factors that contribute to the difference. Spiral galaxies like M31 and M33 have stronger gravitational binding due to their larger mass. This enhanced gravitation promotes more efficient star formation and results in a more pronounced metallicity gradient.
In contrast, irregular galaxies like the LMC and the SMC have smaller mass and weaker gravitational binding, which makes it easier for metal-enriched gas to be expelled by processes such as supernova explosion, stellar wind, and external interaction. Consequently, this leads to weaker and more diffuse metallicity gradients in irregular galaxies.
M31 and M33, being more massive than the LMC and the SMC, are less influenced by external environmental factors.
Their metallicity gradients are primarily shaped by their internal dynamics, with metallicity decreasing gradually from the center to the outskirt.
However, the LMC and the SMC are more susceptible to external disturbances, such as tidal interactions. These factors can perturb or even reshape the metallicity distributions, leading to weaker or less defined metallicity gradients in these irregular galaxies \citep{1976MNRAS.176...31L, 2009A&A...506.1137C, 2009A&A...505..497Y, 2023A&A...677A..91K}.

\begin{figure}
    \centering
    \includegraphics[width=0.9\linewidth]{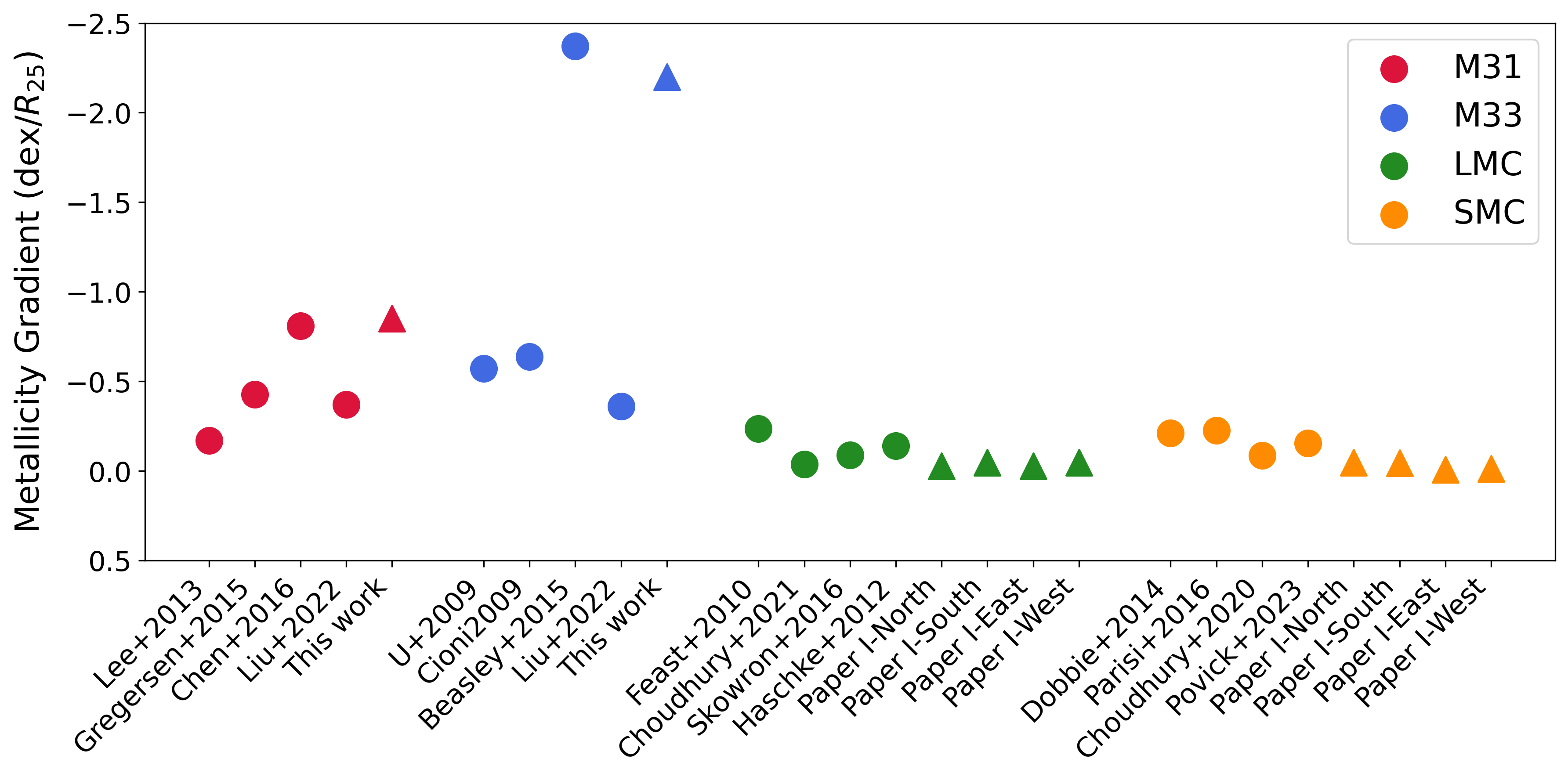}
    \caption{Comparison of metallicity gradients among different galaxies. Circles indicate metallicity gradients from the literature, while triangles represent this work and \citetalias{2024AJ....167..123L}. The metallicity gradient is expressed in units of dex $R_{25}^{-1}$.}
    \label{fig:4-ref-R25}
\end{figure}

In addition, we use the metallicity gradients of these four galaxies to test the statement presented in Section \ref{sec: M33_Z} by \cite{2015MNRAS.448.2030H}: `when metallicity gradients are expressed in units of dex $R^{-1}_{25}$, no correlation is found between metallicity gradient and stellar mass or luminosity'.
Figure \ref{fig:4-ref-R25} presents metallicity gradients compiled from various studies, with each color corresponding to a different galaxy. Circles indicate gradient values reported in the literature, while triangles represent this work and \citetalias{2024AJ....167..123L}. All gradients are expressed in units of dex $R^{-1}_{25}$.
The larger gradient value of the blue triangle may be attributed to the greater dispersion of M33 in our work.
The gradient values shown in the figure are broadly consistent across galaxies and indeed appear to show no clear correlation with galaxy luminosity or stellar mass.

\subsubsection{Extinction Effect} \label{sec:extinction effect}

M31 and M33 own abundant interstellar dust and are subject to the effect of extinction.
Technically, the amount of interstellar extinction for an individual star is not known accurately, therefore, this effect is not accounted for in the data. Still, we perform a comparative analysis between the extinction-corrected samples and the original uncorrected dataset used in this study.
The optical extinction map is taken from \cite{1998ApJ...500..525S} (SFD98) and \cite{2011ApJ...737..103S}.
The SFD98 extinction is derived from the infrared emission that comes not only from the foreground Milky Way but also from M31 or M33 itself, so consequently the derived extinction includes both sources \citep{2021ApJ...912..112W}.
With E(B-V) from SFD98, the extinction in the near-infrared band is calculated with the extinction law by \cite{2019ApJ...877..116W}.
These extinction laws represent the average Galactic extinction and may also be plausible to M31 and M33, as they are spiral galaxies similar to the Milky Way (see, e.g., \cite{1996ApJ...471..203B}). However, it is important to note that the extinction laws specific to M31 and M33 have not yet been well constrained.
M31 spans approximately 4$^{\circ}\!\!.$55 $\times$ 3$^{\circ}\!\!.$23 on the sky, and M33 covers about 1$^{\circ}\!\!.$77 $\times$ 1$^{\circ}\!\!.$73. The SFD98 extinction map has a spatial resolution of 6$^{\prime}\!\!.$1.
To see the change brought about by extinction, two cases are calculated with A$_{\lambda}$=0.5$A_{\lambda}^{\rm SFD98}$ and A$_{\lambda}$=1.0$A_{\lambda}^{\rm SFD98}$ respectively.

The $J-K$ vs. $R_{\rm GC}$ of Sample-half-extinction of M31 within 24 kpc is shown in the bottom panel of Figure \ref{fig: M31_25dis_J-K}, with a step size of 1.5 kpc. An increasing trend is observed within 4.5 kpc, although it is weaker compared to the `no-extinction' data. A prominent bump is visible in the range of 4.5$-$24 kpc, specifically spanning from 10 to 15 kpc, which aligns closely with the locations of the dust rings at 11.5 kpc and 14.5 kpc. This analysis further confirms that regardless of whether extinction is accounted for in the data processing, a clear pattern of elevated $J-K$ values is evident within the 10$-$15 kpc range, highlighting the presence of significant dust structures.

\begin{figure}
    \centering
    \includegraphics[width=0.9\linewidth]{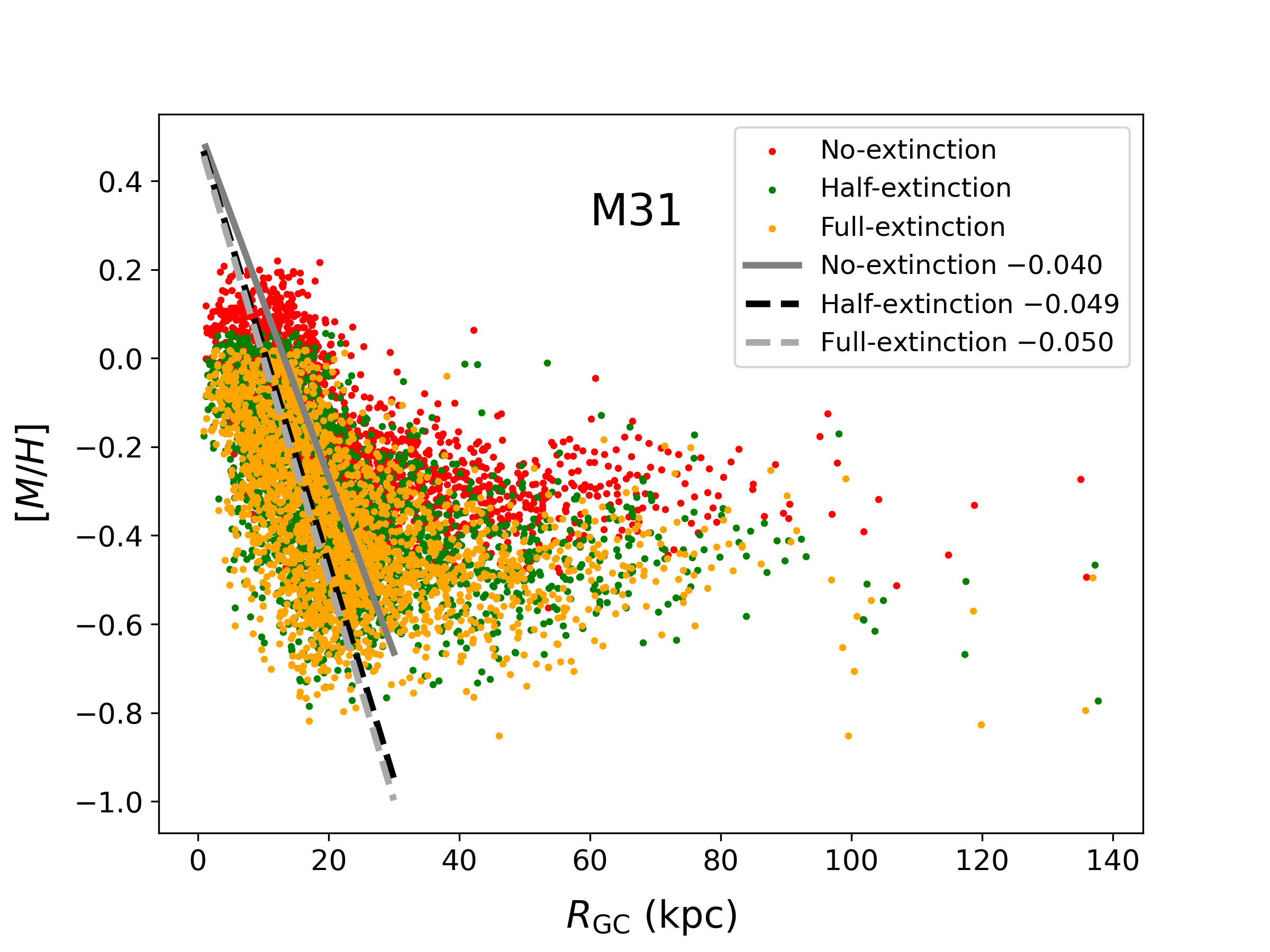}
    \caption{Metallicity distribution in M31 at different extinction corrections. The dots present the [M/H] of each bin.}
    \label{fig:M31-three gradient}
\end{figure}

\begin{figure}
    \centering
    \includegraphics[width=0.7\linewidth]{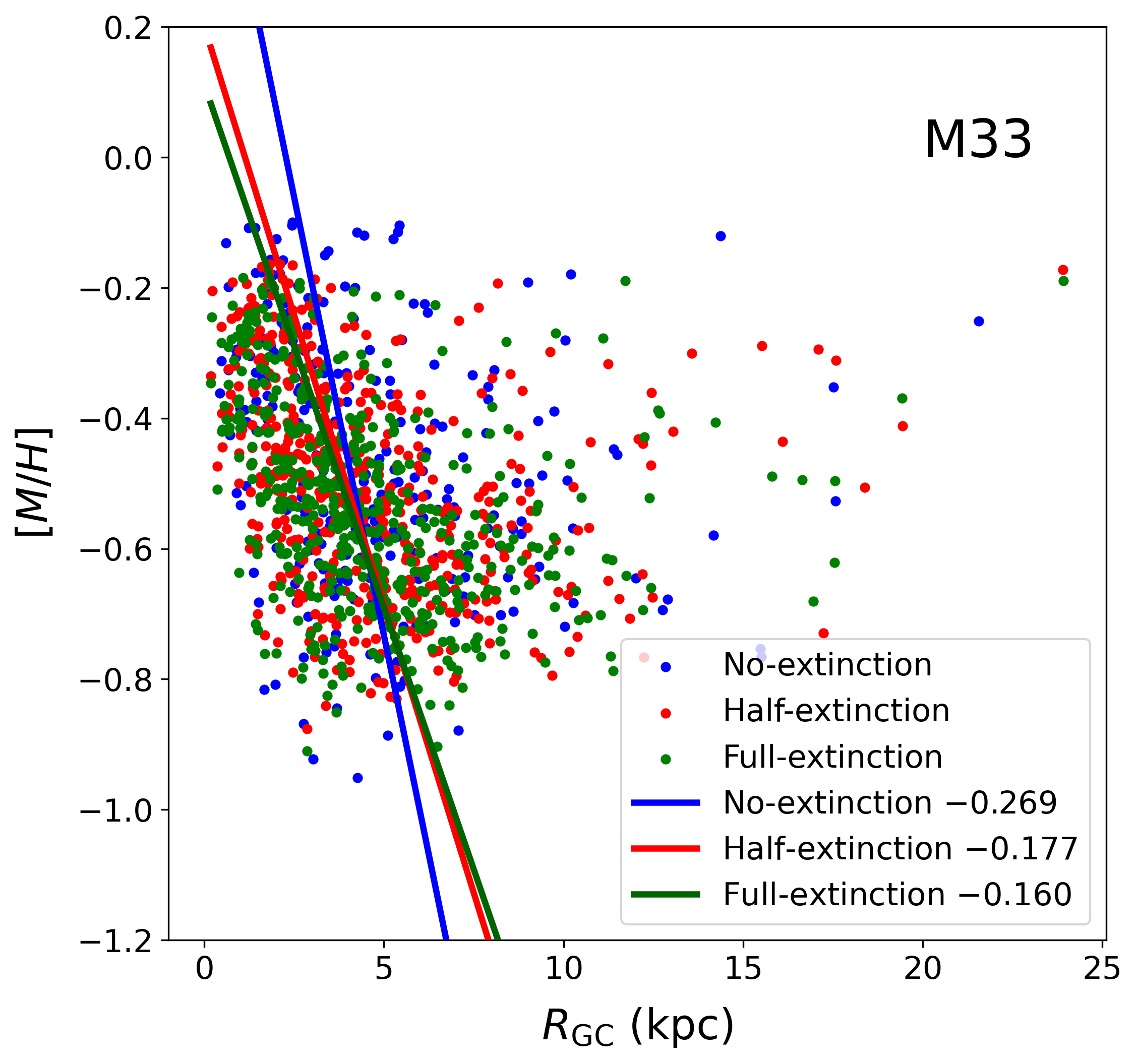}
    \caption{Metallicity distribution in M33 at different extinction corrections. The dots present the [M/H] of each bin.}
    \label{fig:M33-three gradient}
\end{figure}

The metallicity distributions resulted from the different extinction corrections are presented in Figure \ref{fig:M31-three gradient} and Figure \ref{fig:M33-three gradient}. These comparisons reveal that the metallicity gradients derived from the extinction-corrected samples are similar to that with no extinction correction. However, there are notable differences in absolute metallicity.
With no-extinction, the metallicity of M31 increases by 0.1 dex, but it does not change in M33 because the interstellar extinction is very small.

In conclusion, the SFD98 dust map accounts for both foreground and internal extinction. To assess the impact of extinction on our results, we applied different values of $A_{\lambda}$ in the metallicity gradient analysis. The resulting gradients remain consistent across different extinction corrections within the same galaxy, indicating that extinction has no significant effect on the determination of metallicity gradient.

\section{Summary} \label{sec:summary}

Metallicity gradient is a vital tracer of galaxy evolution, revealing how star formation, gas accretion, and mergers redistribute metals and shape a galaxy's chemical structure over time.
Age is considered a potential factor affecting metallicity gradients during galaxy evolution, but its influence has been found to be minor in previous studies \citep{2015AJ....150..189G}. Therefore, we do not include age effects in our analysis.

In order to study the metallicity gradient in M31 and M33, this work uses the member star catalog from \cite{2021ApJ...907...18R}, which excludes foreground dwarf stars with near-infrared color-color diagram as well as the Gaia astrometric measurements.
We use the Voronoi binning to divide the galaxy into multiple sub-regions, the PN method to determine the TRGB position, and the bootstrap method to estimate the error.

The TRGB position in CMD is $J-K$ = 1.195 $\pm$ 0.002 and $K = 17.615 \pm 0.007$ for M31, $J-K$ = 1.100 $\pm$ 0.003 and $K = 18.185 \pm 0.053$ for M33.
A radial gradient in $J-K$ color is observed across both M31 and M33. These color variations are likely indicative of a metallicity gradient, reflecting a decline in chemical abundance from the central regions toward the outskirts of the galaxies.
Overall, the $J-K$ gradient of M31 is $-$0.0055 mag kpc$^{-1}$ within $R_{\rm GC} < 24$ kpc, and $-$0.0002 mag kpc$^{-1}$ with $R_{\rm GC} > 24$ kpc.
A more detailed analysis for $R_{\rm GC}$ in 0$-$24 kpc yields the $J-K$ gradient of 0.003 mag kpc$^{-1}$ for $R_{\rm GC} < 4$ kpc and decreases from 4 to 24 kpc in the way of $J-K=0.299 \times e^{-R_{\rm GC}/33.903}+1.042$.
We identify two structures at 11.5 kpc and 14.5 kpc, which coincide with the dust rings proposed by \citet{2014ApJ...780..172D} and \citet{2014ApJ...786..130L}, and this correspondence is further supported by a comparison with the distribution of RGSs.
The $J-K$ gradient for M33 is analyzed in four directions and fitted by exponential functions. The metallicity gradient generally decreases with $R_{\rm GC}$.
The result for the north direction is $J-K=0.088 \times e^{-R_{\rm GC}/-2.296}+1.113$, for the south direction $J-K=0.112 \times e^{-R_{\rm GC}/3.114}+1.121$, the east direction $J-K=0.192 \times e^{-R_{\rm GC}/-3.113}+1.053$, and the west direction $J-K=1.606 \times e^{-R_{\rm GC}/0.888}+1.114$ ($R_{\rm GC}>2$ kpc).
It is worth noting that there is a positive gradient (0.022 dex kpc$^{-1}$) within 2 kpc in the west direction.

Converting the color index of TRGB into metallicity, the metallicity gradients of M31 and M33 are considered piecewise. For M31, the metallicity gradient is $-0.040 \pm 0.0012$ dex kpc$^{-1}$ within 30 kpc and $-0.001 \pm 0.0002$ dex kpc$^{-1}$ for $R_{\rm GC}$ greater than 30 kpc. For M33, we find that the gradient within 9 kpc is $-$0.269 $\pm$ 0.0206 dex kpc$^{-1}$.
The metallicity gradient of M33 is steeper than that of M31, potentially due to differences in the mass and luminosity of the galaxies. However, the much smaller Pearson correlation between distance and metallicity in M33 suggests that this conclusion requires further consideration.
Our measured metallicity gradient is slightly steeper than those reported in previous studies, consistent with the widely observed trend that metallicity decreases from the centers to the outskirts of spiral galaxies. This negative gradient can be attributed to both spatial variations in star formation efficiency and the timescales of chemical evolution. Central regions, under stronger gravitational potential, tend to retain more gas and dust, forming stars more efficiently and enriching the interstellar medium more rapidly. In contrast, the outer disk, with weaker gravitational confinement and later formation, experiences lower star formation efficiency and has had less time for metal enrichment. These findings align with theoretical models and observational results from both the Milky Way and nearby spiral galaxies \citep{2023A&A...679A..83K, 2021MNRAS.502.5935S, 2010A&A...520A..35M}.

We also compare the metallicity gradients of the LMC, SMC, M31, and M33 to examine the differences between irregular and spiral galaxies. The LMC and SMC have shallower gradients, ranging from $-$0.006 dex kpc$^{-1}$ to $-$0.017 dex kpc$^{-1}$, while M31 and M33 exhibit much steeper gradients ($-$0.040 dex kpc$^{-1}$ and $-$0.269 dex kpc$^{-1}$, respectively). These differences are attributed to stronger gravitational binding in spirals, enhancing star formation, and weaker binding in irregular galaxies, which allows for metal expulsion. External factors like tidal interactions also affect the LMC and SMC. When expressed in units of dex R$^{-1}_{25}$, no clear correlation between metallicity gradient and galaxy luminosity or mass is observed.

M31 and M33 are influenced by interstellar extinction. A comparison between extinction-corrected and uncorrected datasets shows that extinction has little impact on the metallicity gradients. Using the SFD98 dust map, we applied the extinction corrections with zero, half, and full SFD98 values to the data. The results indicate that the metallicity gradients remain consistent across different extinction corrections within each galaxy, suggesting that extinction has a negligible effect on the determination of metallicity gradients, although it does affect the absolute metallicity, i.e. without extinction correction, the metallicity of M31 increases by 0.1 dex.

\acknowledgments{We thank Drs. Jian Gao, Haibo Yuan, Shu Wang, Yuxi Wang, Min Dai, Zhenzhen Shao and Zhetai Cao for their very helpful discussion. We are deeply grateful to the anonymous referee for his/her very helpful suggestions. 
This work is supported by the NSFC projects 12133002, 12203016, and 12203025, CMS-CSST-2021-A09, Shandong Provincial Natural Science Foundation through project ZR2022QA064, Shandong Provincial University Youth Innovation and Technology Support Program through grant No. 2022KJ138.
The numerical computations were conducted on the Qilu Normal University High Performance Computing, Jinan Key Laboratory of Astronomical Data.}

\vspace{5mm}
\facilities{}

\software{Astropy \citep{2013A&A...558A..33A}, TOPCAT \citep{2005ASPC..347...29T}
          }

\bibliography{paper}{}

\begin{thebibliography}{}
\expandafter\ifx\csname natexlab\endcsname\relax\def\natexlab#1{#1}\fi
\providecommand{\url}[1]{\href{#1}{#1}}
\providecommand{\dodoi}[1]{doi:~\href{http://doi.org/#1}{\nolinkurl{#1}}}
\providecommand{\doeprint}[1]{\href{http://ascl.net/#1}{\nolinkurl{http://ascl.net/#1}}}
\providecommand{\doarXiv}[1]{\href{https://arxiv.org/abs/#1}{\nolinkurl{https://arxiv.org/abs/#1}}}

\bibitem[{{Armandroff} \& {Massey}(1991)}]{1991AJ....102..927A}
{Armandroff}, T.~E., \& {Massey}, P. 1991, \aj, 102, 927, \dodoi{10.1086/115924}

\bibitem[{{Astropy Collaboration} {et~al.}(2013){Astropy Collaboration}, {Robitaille}, {Tollerud}, {Greenfield}, {Droettboom}, {Bray}, {Aldcroft}, {Davis}, {Ginsburg}, {Price-Whelan}, {Kerzendorf}, {Conley}, {Crighton}, {Barbary}, {Muna}, {Ferguson}, {Grollier}, {Parikh}, {Nair}, {Unther}, {Deil}, {Woillez}, {Conseil}, {Kramer}, {Turner}, {Singer}, {Fox}, {Weaver}, {Zabalza}, {Edwards}, {Azalee Bostroem}, {Burke}, {Casey}, {Crawford}, {Dencheva}, {Ely}, {Jenness}, {Labrie}, {Lim}, {Pierfederici}, {Pontzen}, {Ptak}, {Refsdal}, {Servillat}, \& {Streicher}}]{2013A&A...558A..33A}
{Astropy Collaboration}, {Robitaille}, T.~P., {Tollerud}, E.~J., {et~al.} 2013, \aap, 558, A33, \dodoi{10.1051/0004-6361/201322068}

\bibitem[{{Balick} {et~al.}(2013){Balick}, {Kwitter}, {Corradi}, \& {Henry}}]{2013ApJ...774....3B}
{Balick}, B., {Kwitter}, K.~B., {Corradi}, R.~L.~M., \& {Henry}, R.~B.~C. 2013, \apj, 774, 3, \dodoi{10.1088/0004-637X/774/1/3}

\bibitem[{{Barmby} {et~al.}(2000){Barmby}, {Huchra}, {Brodie}, {Forbes}, {Schroder}, \& {Grillmair}}]{2000AJ....119..727B}
{Barmby}, P., {Huchra}, J.~P., {Brodie}, J.~P., {et~al.} 2000, \aj, 119, 727, \dodoi{10.1086/301213}

\bibitem[{{Beasley} {et~al.}(2015){Beasley}, {San Roman}, {Gallart}, {Sarajedini}, \& {Aparicio}}]{2015MNRAS.451.3400B}
{Beasley}, M.~A., {San Roman}, I., {Gallart}, C., {Sarajedini}, A., \& {Aparicio}, A. 2015, \mnras, 451, 3400, \dodoi{10.1093/mnras/stv943}

\bibitem[{{Bellazzini} {et~al.}(2004){Bellazzini}, {Ferraro}, {Sollima}, {Pancino}, \& {Origlia}}]{2004A&A...424..199B}
{Bellazzini}, M., {Ferraro}, F.~R., {Sollima}, A., {Pancino}, E., \& {Origlia}, L. 2004, \aap, 424, 199, \dodoi{10.1051/0004-6361:20035910}

\bibitem[{{Bianchi} {et~al.}(1996){Bianchi}, {Clayton}, {Bohlin}, {Hutchings}, \& {Massey}}]{1996ApJ...471..203B}
{Bianchi}, L., {Clayton}, G.~C., {Bohlin}, R.~C., {Hutchings}, J.~B., \& {Massey}, P. 1996, \apj, 471, 203, \dodoi{10.1086/177963}

\bibitem[{{Blair} {et~al.}(1982){Blair}, {Kirshner}, \& {Chevalier}}]{1982ApJ...254...50B}
{Blair}, W.~P., {Kirshner}, R.~P., \& {Chevalier}, R.~A. 1982, \apj, 254, 50, \dodoi{10.1086/159703}

\bibitem[{{Bresolin}(2019)}]{2019MNRAS.488.3826B}
{Bresolin}, F. 2019, \mnras, 488, 3826, \dodoi{10.1093/mnras/stz1947}

\bibitem[{{Bresolin} {et~al.}(2022){Bresolin}, {Kudritzki}, \& {Urbaneja}}]{2022ApJ...940...32B}
{Bresolin}, F., {Kudritzki}, R.-P., \& {Urbaneja}, M.~A. 2022, \apj, 940, 32, \dodoi{10.3847/1538-4357/ac9584}

\bibitem[{{Cappellari} \& {Copin}(2003)}]{2003MNRAS.342..345C}
{Cappellari}, M., \& {Copin}, Y. 2003, \mnras, 342, 345, \dodoi{10.1046/j.1365-8711.2003.06541.x}

\bibitem[{{Chen} {et~al.}(2016){Chen}, {Liu}, {Xiang}, {Yuan}, {Huang}, {Shi}, {Fan}, {Huo}, {Wang}, {Ren}, {Tian}, {Zhang}, {Liu}, {Cao}, {Zhang}, {Hou}, \& {Wang}}]{2016AJ....152...45C}
{Chen}, B., {Liu}, X., {Xiang}, M., {et~al.} 2016, \aj, 152, 45, \dodoi{10.3847/0004-6256/152/2/45}

\bibitem[{{Cioni}(2009)}]{2009A&A...506.1137C}
{Cioni}, M. R.~L. 2009, \aap, 506, 1137, \dodoi{10.1051/0004-6361/200912138}

\bibitem[{{Cioni} {et~al.}(2008){Cioni}, {Irwin}, {Ferguson}, {McConnachie}, {Conn}, {Huxor}, {Ibata}, {Lewis}, \& {Tanvir}}]{2008A&A...487..131C}
{Cioni}, M. R.~L., {Irwin}, M., {Ferguson}, A.~M.~N., {et~al.} 2008, \aap, 487, 131, \dodoi{10.1051/0004-6361:200809366}

\bibitem[{{de Grijs} \& {Bono}(2015)}]{2015AJ....149..179D}
{de Grijs}, R., \& {Bono}, G. 2015, \aj, 149, 179, \dodoi{10.1088/0004-6256/149/6/179}

\bibitem[{{de Grijs} {et~al.}(2014){de Grijs}, {Wicker}, \& {Bono}}]{2014AJ....147..122D}
{de Grijs}, R., {Wicker}, J.~E., \& {Bono}, G. 2014, \aj, 147, 122, \dodoi{10.1088/0004-6256/147/5/122}

\bibitem[{{de Vaucouleurs} {et~al.}(1991){de Vaucouleurs}, {de Vaucouleurs}, {Corwin}, {Buta}, {Paturel}, \& {Fouque}}]{1991rc3..book.....D}
{de Vaucouleurs}, G., {de Vaucouleurs}, A., {Corwin}, Jr., H.~G., {et~al.} 1991, {Third Reference Catalogue of Bright Galaxies}

\bibitem[{{Diehl} \& {Statler}(2006)}]{2006MNRAS.368..497D}
{Diehl}, S., \& {Statler}, T.~S. 2006, \mnras, 368, 497, \dodoi{10.1111/j.1365-2966.2006.10125.x}

\bibitem[{{Draine} {et~al.}(2014){Draine}, {Aniano}, {Krause}, {Groves}, {Sandstrom}, {Braun}, {Leroy}, {Klaas}, {Linz}, {Rix}, {Schinnerer}, {Schmiedeke}, \& {Walter}}]{2014ApJ...780..172D}
{Draine}, B.~T., {Aniano}, G., {Krause}, O., {et~al.} 2014, \apj, 780, 172, \dodoi{10.1088/0004-637X/780/2/172}

\bibitem[{{Durrell} {et~al.}(2001){Durrell}, {Harris}, \& {Pritchet}}]{2001AJ....121.2557D}
{Durrell}, P.~R., {Harris}, W.~E., \& {Pritchet}, C.~J. 2001, \aj, 121, 2557, \dodoi{10.1086/320403}

\bibitem[{{Efron}(1979)}]{Efron1979}
{Efron}, B. 1979, The Annals of Statistics, \dodoi{10.1214/aos/1176344552}

\bibitem[{{Escala} {et~al.}(2021){Escala}, {Gilbert}, {Wojno}, {Kirby}, \& {Guhathakurta}}]{2021AJ....162...45E}
{Escala}, I., {Gilbert}, K.~M., {Wojno}, J., {Kirby}, E.~N., \& {Guhathakurta}, P. 2021, \aj, 162, 45, \dodoi{10.3847/1538-3881/abfec4}

\bibitem[{{Freedman} \& {Madore}(1990)}]{1990ApJ...365..186F}
{Freedman}, W.~L., \& {Madore}, B.~F. 1990, \apj, 365, 186, \dodoi{10.1086/169469}

\bibitem[{{Freedman} {et~al.}(2020){Freedman}, {Madore}, {Hoyt}, {Jang}, {Beaton}, {Lee}, {Monson}, {Neeley}, \& {Rich}}]{2020ApJ...891...57F}
{Freedman}, W.~L., {Madore}, B.~F., {Hoyt}, T., {et~al.} 2020, \apj, 891, 57, \dodoi{10.3847/1538-4357/ab7339}

\bibitem[{{Gibson} {et~al.}(2023){Gibson}, {Zasowski}, {Seth}, {Ashok}, {Goold}, {Wainer}, {Hasselquist}, {Holtzman}, {Imig}, {Bizyaev}, \& {Majewski}}]{2023ApJ...952...23G}
{Gibson}, B.~J., {Zasowski}, G., {Seth}, A., {et~al.} 2023, \apj, 952, 23, \dodoi{10.3847/1538-4357/acd9a9}

\bibitem[{{G{\'o}rski} {et~al.}(2018){G{\'o}rski}, {Pietrzy{\'n}ski}, {Gieren}, {Graczyk}, {Suchomska}, {Karczmarek}, {Cohen}, {Zgirski}, {Wielg{\'o}rski}, {Pilecki}, {Taormina}, {Ko{\l}aczkowski}, \& {Narloch}}]{2018AJ....156..278G}
{G{\'o}rski}, M., {Pietrzy{\'n}ski}, G., {Gieren}, W., {et~al.} 2018, \aj, 156, 278, \dodoi{10.3847/1538-3881/aaeacb}

\bibitem[{{Gregersen} {et~al.}(2015){Gregersen}, {Seth}, {Williams}, {Lang}, {Dalcanton}, {Girardi}, {Skillman}, {Bell}, {Dolphin}, {Fouesneau}, {Guhathakurta}, {Hamren}, {Johnson}, {Kalirai}, {Lewis}, {Monachesi}, \& {Olsen}}]{2015AJ....150..189G}
{Gregersen}, D., {Seth}, A.~C., {Williams}, B.~F., {et~al.} 2015, \aj, 150, 189, \dodoi{10.1088/0004-6256/150/6/189}

\bibitem[{{Grossi} {et~al.}(2008){Grossi}, {Giovanardi}, {Corbelli}, {Giovanelli}, {Haynes}, {Martin}, {Saintonge}, \& {Dowell}}]{2008A&A...487..161G}
{Grossi}, M., {Giovanardi}, C., {Corbelli}, E., {et~al.} 2008, \aap, 487, 161, \dodoi{10.1051/0004-6361:200810220}

\bibitem[{{Ho} {et~al.}(2015){Ho}, {Kudritzki}, {Kewley}, {Zahid}, {Dopita}, {Bresolin}, \& {Rupke}}]{2015MNRAS.448.2030H}
{Ho}, I.~T., {Kudritzki}, R.-P., {Kewley}, L.~J., {et~al.} 2015, \mnras, 448, 2030, \dodoi{10.1093/mnras/stv067}

\bibitem[{{Hoopes} \& {Walterbos}(2000)}]{2000ApJ...541..597H}
{Hoopes}, C.~G., \& {Walterbos}, R. A.~M. 2000, \apj, 541, 597, \dodoi{10.1086/309487}

\bibitem[{{Hoyt} {et~al.}(2024){Hoyt}, {Jang}, {Freedman}, {Madore}, {Lee}, \& {Owens}}]{2024ApJ...975..111H}
{Hoyt}, T.~J., {Jang}, I.~S., {Freedman}, W.~L., {et~al.} 2024, \apj, 975, 111, \dodoi{10.3847/1538-4357/ad7952}

\bibitem[{{Huchra} {et~al.}(1991){Huchra}, {Brodie}, \& {Kent}}]{1991ApJ...370..495H}
{Huchra}, J.~P., {Brodie}, J.~P., \& {Kent}, S.~M. 1991, \apj, 370, 495, \dodoi{10.1086/169836}

\bibitem[{{Ibata} {et~al.}(2014){Ibata}, {Lewis}, {McConnachie}, {Martin}, {Irwin}, {Ferguson}, {Babul}, {Bernard}, {Chapman}, {Collins}, {Fardal}, {Mackey}, {Navarro}, {Pe{\~n}arrubia}, {Rich}, {Tanvir}, \& {Widrow}}]{2014ApJ...780..128I}
{Ibata}, R.~A., {Lewis}, G.~F., {McConnachie}, A.~W., {et~al.} 2014, \apj, 780, 128, \dodoi{10.1088/0004-637X/780/2/128}

\bibitem[{{Ita} {et~al.}(2002){Ita}, {Tanab{\'e}}, {Matsunaga}, {Nakajima}, {Nagashima}, {Nagayama}, {Kato}, {Kurita}, {Nagata}, {Sato}, {Tamura}, {Nakaya}, \& {Nakada}}]{2002MNRAS.337L..31I}
{Ita}, Y., {Tanab{\'e}}, T., {Matsunaga}, N., {et~al.} 2002, \mnras, 337, L31, \dodoi{10.1046/j.1365-8711.2002.06109.x}

\bibitem[{{Jarrett} {et~al.}(2003){Jarrett}, {Chester}, {Cutri}, {Schneider}, \& {Huchra}}]{2003AJ....125..525J}
{Jarrett}, T.~H., {Chester}, T., {Cutri}, R., {Schneider}, S.~E., \& {Huchra}, J.~P. 2003, \aj, 125, 525, \dodoi{10.1086/345794}

\bibitem[{{Kang} {et~al.}(2023){Kang}, {Kudritzki}, \& {Zhang}}]{2023A&A...679A..83K}
{Kang}, X., {Kudritzki}, R.-P., \& {Zhang}, F. 2023, \aap, 679, A83, \dodoi{10.1051/0004-6361/202347677}

\bibitem[{{Khoperskov} {et~al.}(2023){Khoperskov}, {Minchev}, {Libeskind}, {Belokurov}, {Steinmetz}, {Gomez}, {Grand}, {Hoffman}, {Knebe}, {Sorce}, {Spaare}, {Tempel}, \& {Vogelsberger}}]{2023A&A...677A..91K}
{Khoperskov}, S., {Minchev}, I., {Libeskind}, N., {et~al.} 2023, \aap, 677, A91, \dodoi{10.1051/0004-6361/202244234}

\bibitem[{{Kwitter} {et~al.}(2012){Kwitter}, {Lehman}, {Balick}, \& {Henry}}]{2012ApJ...753...12K}
{Kwitter}, K.~B., {Lehman}, E. M.~M., {Balick}, B., \& {Henry}, R.~B.~C. 2012, \apj, 753, 12, \dodoi{10.1088/0004-637X/753/1/12}

\bibitem[{{Larson}(1976)}]{1976MNRAS.176...31L}
{Larson}, R.~B. 1976, \mnras, 176, 31, \dodoi{10.1093/mnras/176.1.31}

\bibitem[{{Lee} {et~al.}(2013){Lee}, {Kodric}, {Seitz}, {Riffeser}, {Koppenhoefer}, {Bender}, {Hopp}, {G{\"o}ssl}, {Snigula}, {Burgett}, {Chambers}, {Flewelling}, {Hodapp}, {Kaiser}, {Kudritzki}, {Price}, {Tonry}, \& {Wainscoat}}]{2013ApJ...777...35L}
{Lee}, C.~H., {Kodric}, M., {Seitz}, S., {et~al.} 2013, \apj, 777, 35, \dodoi{10.1088/0004-637X/777/1/35}

\bibitem[{{Lee} \& {Lee}(2014)}]{2014ApJ...786..130L}
{Lee}, J.~H., \& {Lee}, M.~G. 2014, \apj, 786, 130, \dodoi{10.1088/0004-637X/786/2/130}

\bibitem[{{Lee} {et~al.}(1993){Lee}, {Freedman}, \& {Madore}}]{1993ApJ...417..553L}
{Lee}, M.~G., {Freedman}, W.~L., \& {Madore}, B.~F. 1993, \apj, 417, 553, \dodoi{10.1086/173334}

\bibitem[{{Li} {et~al.}(2024){Li}, {Jiang}, \& {Ren}}]{2024AJ....167..123L}
{Li}, Y., {Jiang}, B., \& {Ren}, Y. 2024, \aj, 167, 123, \dodoi{10.3847/1538-3881/ad23e8}

\bibitem[{{Liu} {et~al.}(2022){Liu}, {Kudritzki}, {Zhao}, {Urbaneja}, {Huang}, {Zhang}, \& {Zhao}}]{2022ApJ...932...29L}
{Liu}, C., {Kudritzki}, R.-P., {Zhao}, G., {et~al.} 2022, \apj, 932, 29, \dodoi{10.3847/1538-4357/ac69cc}

\bibitem[{{Ma}(2001)}]{2001ChPhL..18.1420M}
{Ma}, J. 2001, Chinese Physics Letters, 18, 1420, \dodoi{10.1088/0256-307X/18/10/339}

\bibitem[{{MacArthur} {et~al.}(2004){MacArthur}, {Courteau}, {Bell}, \& {Holtzman}}]{2004ApJS..152..175M}
{MacArthur}, L.~A., {Courteau}, S., {Bell}, E., \& {Holtzman}, J.~A. 2004, \apjs, 152, 175, \dodoi{10.1086/383525}

\bibitem[{{Magrini} {et~al.}(2007){Magrini}, {Corbelli}, \& {Galli}}]{2007A&A...470..843M}
{Magrini}, L., {Corbelli}, E., \& {Galli}, D. 2007, \aap, 470, 843, \dodoi{10.1051/0004-6361:20077215}

\bibitem[{{Magrini} {et~al.}(2009){Magrini}, {Stanghellini}, \& {Villaver}}]{2009ApJ...696..729M}
{Magrini}, L., {Stanghellini}, L., \& {Villaver}, E. 2009, \apj, 696, 729, \dodoi{10.1088/0004-637X/696/1/729}

\bibitem[{{Marcon-Uchida} {et~al.}(2010){Marcon-Uchida}, {Matteucci}, \& {Costa}}]{2010A&A...520A..35M}
{Marcon-Uchida}, M.~M., {Matteucci}, F., \& {Costa}, R.~D.~D. 2010, \aap, 520, A35, \dodoi{10.1051/0004-6361/200913933}

\bibitem[{{Massey} {et~al.}(2021){Massey}, {Neugent}, {Levesque}, {Drout}, \& {Courteau}}]{2021AJ....161...79M}
{Massey}, P., {Neugent}, K.~F., {Levesque}, E.~M., {Drout}, M.~R., \& {Courteau}, S. 2021, \aj, 161, 79, \dodoi{10.3847/1538-3881/abd01f}

\bibitem[{{McConnachie} {et~al.}(2004){McConnachie}, {Irwin}, {Ferguson}, {Ibata}, {Lewis}, \& {Tanvir}}]{2004MNRAS.350..243M}
{McConnachie}, A.~W., {Irwin}, M.~J., {Ferguson}, A.~M.~N., {et~al.} 2004, \mnras, 350, 243, \dodoi{10.1111/j.1365-2966.2004.07637.x}

\bibitem[{{Melchior} \& {Combes}(2011)}]{2011A&A...536A..52M}
{Melchior}, A.~L., \& {Combes}, F. 2011, \aap, 536, A52, \dodoi{10.1051/0004-6361/201016031}

\bibitem[{{M{\'e}ndez} {et~al.}(2002){M{\'e}ndez}, {Davis}, {Moustakas}, {Newman}, {Madore}, \& {Freedman}}]{2002AJ....124..213M}
{M{\'e}ndez}, B., {Davis}, M., {Moustakas}, J., {et~al.} 2002, \aj, 124, 213, \dodoi{10.1086/341168}

\bibitem[{{Mould} \& {Kristian}(1986)}]{1986ApJ...305..591M}
{Mould}, J., \& {Kristian}, J. 1986, \apj, 305, 591, \dodoi{10.1086/164273}

\bibitem[{{Ogami} {et~al.}(2024){Ogami}, {Tanaka}, {Komiyama}, {Chiba}, {Guhathakurta}, {Kirby}, {Wyse}, {Filion}, {Gilbert}, {Escala}, {Mori}, {Kirihara}, {Ishigaki}, {Hayashi}, {Lee}, {Sharma}, {Kalirai}, \& {Lupton}}]{2024MNRAS.tmp.2458O}
{Ogami}, I., {Tanaka}, M., {Komiyama}, Y., {et~al.} 2024, \mnras, \dodoi{10.1093/mnras/stae2527}

\bibitem[{{Pawlak}(2021)}]{2021A&A...649A.110P}
{Pawlak}, M. 2021, \aap, 649, A110, \dodoi{10.1051/0004-6361/202038642}

\bibitem[{{Rana}(1991)}]{1991ARA&A..29..129R}
{Rana}, N.~C. 1991, \araa, 29, 129, \dodoi{10.1146/annurev.aa.29.090191.001021}

\bibitem[{{Ren} {et~al.}(2021{\natexlab{a}}){Ren}, {Jiang}, {Yang}, {Wang}, {Jian}, \& {Ren}}]{2021ApJ...907...18R}
{Ren}, Y., {Jiang}, B., {Yang}, M., {et~al.} 2021{\natexlab{a}}, \apj, 907, 18, \dodoi{10.3847/1538-4357/abcda5}

\bibitem[{{Ren} {et~al.}(2021{\natexlab{b}}){Ren}, {Jiang}, {Yang}, {Wang}, \& {Ren}}]{2021ApJ...923..232R}
{Ren}, Y., {Jiang}, B., {Yang}, M., {Wang}, T., \& {Ren}, T. 2021{\natexlab{b}}, \apj, 923, 232, \dodoi{10.3847/1538-4357/ac307b}

\bibitem[{{Saglia} {et~al.}(2018){Saglia}, {Opitsch}, {Fabricius}, {Bender}, {Bla{\~n}a}, \& {Gerhard}}]{2018A&A...618A.156S}
{Saglia}, R.~P., {Opitsch}, M., {Fabricius}, M.~H., {et~al.} 2018, \aap, 618, A156, \dodoi{10.1051/0004-6361/201732517}

\bibitem[{{Sakai} {et~al.}(1996){Sakai}, {Madore}, \& {Freedman}}]{1996ApJ...461..713S}
{Sakai}, S., {Madore}, B.~F., \& {Freedman}, W.~L. 1996, \apj, 461, 713, \dodoi{10.1086/177096}

\bibitem[{{Salaris} \& {Cassisi}(2005)}]{2005essp.book.....S}
{Salaris}, M., \& {Cassisi}, S. 2005, {Evolution of Stars and Stellar Populations}

\bibitem[{{Salaris} \& {Girardi}(2005)}]{2005MNRAS.357..669S}
{Salaris}, M., \& {Girardi}, L. 2005, \mnras, 357, 669, \dodoi{10.1111/j.1365-2966.2005.08689.x}

\bibitem[{{S{\'a}nchez-Bl{\'a}zquez} {et~al.}(2014){S{\'a}nchez-Bl{\'a}zquez}, {Rosales-Ortega}, {M{\'e}ndez-Abreu}, {P{\'e}rez}, {S{\'a}nchez}, {Zibetti}, {Aguerri}, {Bland-Hawthorn}, {Catal{\'a}n-Torrecilla}, {Cid Fernandes}, {de Amorim}, {de Lorenzo-Caceres}, {Falc{\'o}n-Barroso}, {Galazzi}, {Garc{\'\i}a Benito}, {Gil de Paz}, {Gonz{\'a}lez Delgado}, {Husemann}, {Iglesias-P{\'a}ramo}, {Jungwiert}, {Marino}, {M{\'a}rquez}, {Mast}, {Mendoza}, {Moll{\'a}}, {Papaderos}, {Ruiz-Lara}, {van de Ven}, {Walcher}, \& {Wisotzki}}]{2014A&A...570A...6S}
{S{\'a}nchez-Bl{\'a}zquez}, P., {Rosales-Ortega}, F.~F., {M{\'e}ndez-Abreu}, J., {et~al.} 2014, \aap, 570, A6, \dodoi{10.1051/0004-6361/201423635}

\bibitem[{{Schlafly} \& {Finkbeiner}(2011)}]{2011ApJ...737..103S}
{Schlafly}, E.~F., \& {Finkbeiner}, D.~P. 2011, \apj, 737, 103, \dodoi{10.1088/0004-637X/737/2/103}

\bibitem[{{Schlegel} {et~al.}(1998){Schlegel}, {Finkbeiner}, \& {Davis}}]{1998ApJ...500..525S}
{Schlegel}, D.~J., {Finkbeiner}, D.~P., \& {Davis}, M. 1998, \apj, 500, 525, \dodoi{10.1086/305772}

\bibitem[{{Searle}(1971)}]{1971ApJ...168..327S}
{Searle}, L. 1971, \apj, 168, 327, \dodoi{10.1086/151090}

\bibitem[{{Shao} {et~al.}(2025){Shao}, {Wang}, {Jiang}, {Wang}, {Ge}, \& {Zhu}}]{2025ApJ...980..218S}
{Shao}, Z., {Wang}, S., {Jiang}, B., {et~al.} 2025, \apj, 980, 218, \dodoi{10.3847/1538-4357/adae8e}

\bibitem[{{Sharda} {et~al.}(2021){Sharda}, {Krumholz}, {Wisnioski}, {Forbes}, {Federrath}, \& {Acharyya}}]{2021MNRAS.502.5935S}
{Sharda}, P., {Krumholz}, M.~R., {Wisnioski}, E., {et~al.} 2021, \mnras, 502, 5935, \dodoi{10.1093/mnras/stab252}

\bibitem[{{Smith} {et~al.}(2012){Smith}, {Eales}, {Gomez}, {Roman-Duval}, {Fritz}, {Braun}, {Baes}, {Bendo}, {Blommaert}, {Boquien}, {Boselli}, {Clements}, {Cooray}, {Cortese}, {De Looze}, {Ford}, {Gear}, {Gentile}, {Gordon}, {Kirk}, {Lebouteiller}, {Madden}, {Mentuch}, {O'Halloran}, {Page}, {Schulz}, {Spinoglio}, {Verstappen}, {Wilson}, \& {Thilker}}]{2012ApJ...756...40S}
{Smith}, M.~W.~L., {Eales}, S.~A., {Gomez}, H.~L., {et~al.} 2012, \apj, 756, 40, \dodoi{10.1088/0004-637X/756/1/40}

\bibitem[{{Taylor}(2005)}]{2005ASPC..347...29T}
{Taylor}, M.~B. 2005, in Astronomical Society of the Pacific Conference Series, Vol. 347, Astronomical Data Analysis Software and Systems XIV, ed. P.~{Shopbell}, M.~{Britton}, \& R.~{Ebert}, 29

\bibitem[{{U} {et~al.}(2009){U}, {Urbaneja}, {Kudritzki}, {Jacobs}, {Bresolin}, \& {Przybilla}}]{2009ApJ...704.1120U}
{U}, V., {Urbaneja}, M.~A., {Kudritzki}, R.-P., {et~al.} 2009, \apj, 704, 1120, \dodoi{10.1088/0004-637X/704/2/1120}

\bibitem[{{Valenti} {et~al.}(2004){Valenti}, {Ferraro}, \& {Origlia}}]{2004MNRAS.354..815V}
{Valenti}, E., {Ferraro}, F.~R., \& {Origlia}, L. 2004, \mnras, 354, 815, \dodoi{10.1111/j.1365-2966.2004.08249.x}

\bibitem[{{Wang} \& {Chen}(2019)}]{2019ApJ...877..116W}
{Wang}, S., \& {Chen}, X. 2019, \apj, 877, 116, \dodoi{10.3847/1538-4357/ab1c61}

\bibitem[{{Wang} {et~al.}(2021){Wang}, {Jiang}, {Ren}, {Yang}, \& {Li}}]{2021ApJ...912..112W}
{Wang}, T., {Jiang}, B., {Ren}, Y., {Yang}, M., \& {Li}, J. 2021, \apj, 912, 112, \dodoi{10.3847/1538-4357/abed4b}

\bibitem[{{Williams}(2003)}]{2003AJ....126.1312W}
{Williams}, B.~F. 2003, \aj, 126, 1312, \dodoi{10.1086/377347}

\bibitem[{{Williams} {et~al.}(2009){Williams}, {Dalcanton}, {Dolphin}, {Holtzman}, \& {Sarajedini}}]{2009ApJ...695L..15W}
{Williams}, B.~F., {Dalcanton}, J.~J., {Dolphin}, A.~E., {Holtzman}, J., \& {Sarajedini}, A. 2009, \apjl, 695, L15, \dodoi{10.1088/0004-637X/695/1/L15}

\bibitem[{{Williams} {et~al.}(2015){Williams}, {Lang}, {Dalcanton}, {Dolphin}, {Weisz}, {Bell}, {Bianchi}, {Byler}, {Gilbert}, {Girardi}, {Gordon}, {Gregersen}, {Johnson}, {Kalirai}, {Lauer}, {Monachesi}, {Rosenfield}, {Seth}, \& {Skillman}}]{2015yCat..22150009W}
{Williams}, B.~F., {Lang}, D., {Dalcanton}, J.~J., {et~al.} 2015, {VizieR Online Data Catalog: PHAT X. UV-IR photometry of M31 stars (Williams+, 2014)}, VizieR On-line Data Catalog: J/ApJS/215/9. Originally published in: 2014ApJS..215....9W, \dodoi{10.26093/cds/vizier.22150009}

\bibitem[{{Yin} {et~al.}(2009){Yin}, {Hou}, {Prantzos}, {Boissier}, {Chang}, {Shen}, \& {Zhang}}]{2009A&A...505..497Y}
{Yin}, J., {Hou}, J.~L., {Prantzos}, N., {et~al.} 2009, \aap, 505, 497, \dodoi{10.1051/0004-6361/200912316}

\bibitem[{{Zaritsky} {et~al.}(1994){Zaritsky}, {Kennicutt}, \& {Huchra}}]{1994ApJ...420...87Z}
{Zaritsky}, D., {Kennicutt}, Jr., R.~C., \& {Huchra}, J.~P. 1994, \apj, 420, 87, \dodoi{10.1086/173544}

\bibitem[{{Zurita} \& {Bresolin}(2012)}]{2012MNRAS.427.1463Z}
{Zurita}, A., \& {Bresolin}, F. 2012, \mnras, 427, 1463, \dodoi{10.1111/j.1365-2966.2012.22075.x}

\end{thebibliography}
\bibliographystyle{aasjournal}

\end{CJK*}
\end{document}